\documentclass[11pt,a4paper]{article}
\pdfoutput=1 

\usepackage{jheppub} 

\usepackage[T1]{fontenc} 
\usepackage{slashed}
\usepackage[normalem]{ulem}
\usepackage{xcolor}
\usepackage{tikz}
\usepackage{mathtools}
\usetikzlibrary{decorations.pathmorphing}
\tikzset{snake it/.style={decorate, decoration=snake}}

\def\spa#1.#2{\left\langle#1\,#2\right\rangle}
\def\spb#1.#2{\left[#1\,#2\right]}
\def\spash#1.#2{\spa{\smash{#1}}.{\smash{#2}}}
\def\spbsh#1.#2{\spb{\smash{#1}}.{\smash{#2}}}
\def\sand#1.#2.#3{%
\left\langle\smash{#1}{\vphantom1}^{-}\right|{#2}%
\left|\smash{#3}{\vphantom1}^{-}\right\rangle}
\def\sandpp#1.#2.#3{%
\left\langle\smash{#1}{\vphantom1}^{+}\right|{#2}%
\left|\smash{#3}{\vphantom1}^{+}\right\rangle}
\def\sandpm#1.#2.#3{%
\left\langle\smash{#1}{\vphantom1}^{+}\right|{#2}%
\left|\smash{#3}{\vphantom1}^{-}\right\rangle}
\def\sandmp#1.#2.#3{%
\left\langle\smash{#1}{\vphantom1}^{-}\right|{#2}%
\left|\smash{#3}{\vphantom1}^{+}\right\rangle}

\def\nn{\nonumber}

\def\eqn#1{eq.~(\ref{#1})}

\def\be{\begin{equation}}
\def\ee{\end{equation}}
\def\bea{\begin{eqnarray}}
\def\eea{\end{eqnarray}}
\def\ba{\begin{eqnarray}}
\def\ea{\end{eqnarray}}

\def\HYPER{\Phi}

\newcommand{\bop}{\vartheta}

\title{Off-Shell\,Color-Kinematics\,Duality\,for\,Chern-Simons}

\author[a]{Maor Ben-Shahar}
\author[a,b]{and Henrik Johansson}

\affiliation[a]{Department of Physics and Astronomy, Uppsala University, \\ Box 516, 75120 Uppsala, Sweden}
\affiliation[b]{Nordita, Stockholm University and KTH Royal Institute of Technology, \\ Hannes Alfv\'{e}ns v\"{a}g 12, 10691 Stockholm, Sweden}

\emailAdd{benshahar.maor@physics.uu.se}
\emailAdd{henrik.johansson@physics.uu.se}

\abstract{Many gauge theories possess a hidden duality between color and kinematics in
their on-shell scattering amplitudes. An open problem is to formulate an off-shell realization
of the duality, thus manifesting a kinematic algebra. We show that 3D~Chern-Simons~(CS)
theory in Lorenz gauge obeys off-shell color-kinematics duality. This holds both for the
gauge field and the BRST ghosts, and the duality is manifest in the Feynman rules. A
kinematic algebra can be formulated through a second-order differential operator (Poisson
bracket) acting on the off-shell fields, and it corresponds to 3D volume-preserving diffeomorphisms, generated by functions in Lorenz gauge.
We consider several admissible double-copy constructions of CS theory with Yang-Mills theory, a higher-derivative $(DF)^2$ gauge theory, or CS theory itself.
To obtain non-vanishing amplitudes, we deform pure CS theory by including
the maximum amount of adjoint matter that respects the on-shell duality. This gives a new formulation of an ${\cal N}=4$ CS-matter theory, with fields of unusual statistics. We argue that the color-stripped tree amplitudes of this theory are equivalent to those of the Gaiotto-Witten ${\cal N}=4$ CS theory with bi-fundamental matter. We further show that the double copy of the ${\cal N}=4$ CS theory with itself corresponds to maximally supersymmetric ${\cal N}=8$ Dirac-Born-Infeld theory. 
}

\preprint{ UUITP-63/21 
\\ \phantom{~} \hfill NORDITA 2021-152
}  

\begin{document} 
\maketitle
\flushbottom

\newpage

\section{Introduction}
Non-abelian gauge theories have provided a rich source of remarkable physical and mathematical insights. One revelation that emerged relatively recently is the Bern-Carrasco-Johansson (BCJ) duality between color and kinematics~\citep{Bern:2008qj,Bern:2010ue,Bern:2019prr}. The duality states that quantum scattering amplitudes in many of our most cherished gauge theories may be organized through cubic diagrams that exhibit two dual structures: color factors and kinematic numerators that obey the same Lie-algebraic relations. 
The color factors inherit their relations from the underlying Jacobi identity and Lie algebra of the gauge group, whereas the algebraic origin of the kinematic relations is still shrouded in mystery.

For gauge theories with adjoint massless fields, the existence of color-kinematics duality at tree level is equivalent to the existence of BCJ amplitude relations~\cite{Bern:2008qj,Stieberger:2009hq,Bjerrum-Bohr:2009ulz,Feng:2010my, Chiodaroli:2017ngp}. The color-kinematics duality and the BCJ relations were first formulated for tree-level pure Yang-Mills theory~\cite{Bern:2008qj} and later the duality was generalized to loop-level amplitudes~\cite{Bern:2010ue}, including supersymmetry. The first complete derivations of the tree-level BCJ relations came from string theory~\cite{Stieberger:2009hq,Bjerrum-Bohr:2009ulz} and on-shell recursion~\cite{Feng:2010my} which implied the extension to many supersymmetric purely-adjoint theories. Later work discovered that many different types of gauge theories exhibit the same duality~\cite{Bargheer:2012gv,Huang:2012wr,Broedel:2012rc,Chiodaroli:2014xia,Chiodaroli:2015rdg,Johansson:2017srf,Chiodaroli:2018dbu,Johansson:2018ues,Johansson:2019dnu,Bautista:2019evw,Plefka:2019wyg}, including theories with fundamental matter~\cite{Johansson:2014zca,Johansson:2015oia,Chiodaroli:2013upa}, as well as scalar effective field theories~\cite{Chen:2013fya,Cheung:2016prv,Carrasco:2016ldy,Mafra:2016mcc,Carrasco:2016ygv,Low:2019wuv,Cheung:2020qxc,Rodina:2021isd, Chi:2021mio} without gauge fields.
Further non-trivial calculations at loop level established the duality more firmly~\cite{Bern:2010ue,Carrasco:2011mn,Bern:2012uf,Boels:2013bi,Bjerrum-Bohr:2013iza,Bern:2013yya,Nohle:2013bfa,Mogull:2015adi,Mafra:2015mja,He:2015wgf,Johansson:2017bfl,Hohenegger:2017kqy,Mafra:2017ioj,Faller:2018vdz,Kalin:2018thp,Duhr:2019ywc,Geyer:2019hnn,Edison:2020uzf,Casali:2020knc,DHoker:2020prr,Carrasco:2020ywq,Bridges:2021ebs}, and later included extensions to loop-level form factors~\cite{Boels:2012ew,Yang:2016ear,Boels:2017ftb,Lin:2020dyj,Lin:2021qol,Lin:2021pne,Lin:2021lqo}. Recently the color-kinematics duality has been extended to curved-space amplitudes and correlators~\cite{Adamo:2017nia,Farrow:2018yni,Adamo:2018mpq,Lipstein:2019mpu,Prabhu:2020avf, Casali:2020vuy, Armstrong:2020woi,Albayrak:2020fyp,Adamo:2020qru,Alday:2021odx,Diwakar:2021juk,Zhou:2021gnu,Sivaramakrishnan:2021srm, Adamo:2021rfq}. 

A central corollary to the color-kinematics duality is the double-copy construction: once numerators obeying the duality are found, they can be used to replace color factors in gauge theory amplitudes, thus obtaining gravity amplitudes \citep{Bern:2008qj,Bern:2010ue}. For tree-level amplitudes in purely-adjoint gauge theories, the application of the double copy is equivalent to the field-theory limit of the Kawai-Lewellen-Tye (KLT) relations~\cite{Kawai:1985xq} between open and closed string amplitudes. The double-copy applies to a large web of gravitational theories~\cite{Bern:2010ue,Bern:2010yg,Bern:2011rj,Johansson:2017bfl,Boucher-Veronneau:2011rlc,Bern:2013uka,Bern:2014sna,Chiodaroli:2015wal,Chiodaroli:2017ehv,Bern:2018jmv,Ben-Shahar:2018uie,Bern:2021ppb,Chiodaroli:2021eug,Carrasco:2021bmu,Brandhuber:2021eyq, Diaz-Jaramillo:2021wtl, Cho:2021nim, Adamo:2021dfg, Witzany:2021dru, Li:2021yfk, Yan:2021hte, Menezes:2021dyp}, including string theory~\cite{Mafra:2011nw,Broedel:2013tta,Stieberger:2014hba,Huang:2016tag,Azevedo:2018dgo,Geyer:2021oox, Edison:2021ebi} and  ambi-twistor string constructions~\cite{Cachazo:2012da,Mason:2013sva,Geyer:2014fka, Casali:2015vta,Geyer:2015bja,Geyer:2016wjx}. The double copy has been extended to classical solutions in general relativity, such as Schwarzschild and Kerr black holes~\cite{Monteiro:2014cda,Luna:2015paa,Luna:2016hge,Bahjat-Abbas:2017htu,Carrillo-Gonzalez:2017iyj,Berman:2018hwd,CarrilloGonzalez:2019gof,Goldberger:2019xef,Huang:2019cja,Alawadhi:2019urr,Bahjat-Abbas:2020cyb,Easson:2020esh,Emond:2020lwi, Godazgar:2020zbv,Chacon:2021wbr,Chacon:2020fmr,Alfonsi:2020lub, Monteiro:2020plf, White:2020sfn, Elor:2020nqe,Pasarin:2020qoa,Gonzo:2021drq, Lazauskas:2021vcg,Godazgar:2021iae,Easson:2021asd,Monteiro:2021ztt} and calculations relevant to gravitational-wave physics~\cite{Luna:2016due,Goldberger:2016iau,Luna:2017dtq,Shen:2018ebu,Plefka:2018dpa,Bern:2019nnu,Plefka:2019hmz,Bern:2019crd,Bern:2020buy,Almeida:2020mrg,Bern:2021dqo}.

The topic of duality-satisfying numerators and BCJ amplitude relations in massless gauge theories has by now a vast literature. Amplitude relations have been studied from many complementary perspectives, such as string theory, scattering equations and positive geometry~\cite{Stieberger:2009hq,Bjerrum-Bohr:2009ulz,Cachazo:2012uq,Arkani-Hamed:2017mur,Mizera:2019blq}.  By now there exist many direct methods for the construction of BCJ numerators~\cite{Bjerrum-Bohr:2010pnr,Mafra:2011kj,Fu:2012uy,Mafra:2015vca,Bjerrum-Bohr:2016axv,Du:2017kpo,Chen:2017bug,Fu:2018hpu,Edison:2020ehu,He:2021lro,Bridges:2021ebs, Ahmadiniaz:2021fey,Cheung:2021zvb,Ahmadiniaz:2021ayd,Brandhuber:2021bsf}. However, to gain a complete understanding of this topic, direct exploration of the structure of the underlying kinematic algebra is necessary.  
 
The first construction of a kinematic algebra was formulated for self-dual Yang-Mills theory by Monteiro and O'Connell \cite{Monteiro:2011pc}. They showed that the Feynman rules of the theory can be obtained from nested commutators of generators of area-preserving diffeomorphisms. While self-dual Yang-Mills theory has vanishing tree amplitudes the one-loop amplitudes are non-zero~\cite{Cangemi:1996rx}; their color-kinematics properties were explored in ref.~\cite{Boels:2013bi}.
The second kinematic algebra to be constructed is by Cheung and Shen \cite{Cheung:2016prv}, who realized a cubic Lagrangian for the $SU(N_c)$ non-linear
sigma model (NLSM), with Feynman rules obeying color-kinematics
duality at tree and one-loop level. In refs. \cite{Chen:2019ywi,Chen:2021chy} a kinematic algebra was formulated for Yang-Mills theory up to the next-to-MHV level, where the maximally-helicity-violating (MHV) sector was observed to be unique for local BCJ numerators, and secretly given by the Cheung and Shen Lagrangian~\cite{Cheung:2016prv}. Interestingly, the same color-kinematics-satisfying Lagrangian was later used in a different context to give a non-abelian generalization of the Navier-Stokes equations~\cite{Cheung:2020djz,Keeler:2020rcv}. In the past few months, two new results have made significant progress in elucidating the kinematical algebra of Yang Mills theory. In ref.~\cite{Cheung:2021zvb} a covariant version of color-kinematics duality was proposed and it enabled double-copy manipulations at the Lagrangian level and gave explicit all-multiplicity BCJ numerators.  In ref.~\cite{Brandhuber:2021bsf} a kinematical Hopf algebra was introduced that had formal generators that mapped to explicit BCJ numerators, both in pure Yang-Mills theory and when coupling heavy-mass particles to Yang-Mills theory. 

Other approaches towards elucidating the underlying structure of color-kinematics duality include explicit brute-force constructions of Yang-Mills Lagrangians with manifest color-kinematics duality up sixth order in the fields~\cite{Bern:2010yg,Tolotti:2013caa}. See also refs.~\cite{Ferrero:2020vww,Beneke:2021ilf} on Lagrangian double copies up to cubic order. Alternatively, there are approaches that construct Berends-Giele currents in so-called ``BCJ gauge''~\cite{Lee:2015upy,Bridges:2019siz}, as well as other advanced mathematical explorations using recursive structures, homotopy algebras and diffeomorphism symmetry~\cite{Fu:2016plh,Lopez-Arcos:2019hvg,Reiterer:2019dys,Gomez:2020vat, Fu:2020frx, Borsten:2019prq, Borsten:2020zgj,Borsten:2020xbt,Borsten:2021hua,Borsten:2021zir,Borsten:2021rmh, Frenkel:2020djn, Campiglia:2021srh}.

In this paper we explore the algebraic origins of the color-kinematics duality by studying a simpler gauge theory, namely 3D Chern-Simons theory. In particular, here we uncover that pure Chern-Simons theory admits an off-shell formulation of color-kinematics duality, and that an explicit kinematic algebra can be written down in terms of generators and structure constants of a volume-preserving diffeomorphism algebra. Furthermore, we extend the web of double-copy theories to include several Chern-Simons-matter theories that have not been previously considered in the context of color-kinematics duality. Our results rely on using the vanilla formulation~\cite{Bern:2008qj} of color-kinematics duality and double copy, that expands the adjoint-representation amplitude in cubic massless diagrams. Thus our work is orthogonal to previous more exotic color-kinematics constructions in 3D.
 
Almost a decade ago it was shown~\cite{Bargheer:2012gv,Huang:2012wr,Huang:2013kca} that maximally supersymmetric Chern-Simons theory, ${\cal N}=8$ Bagger-Lambert-Gustavsson (BLG) theory~\cite{Bagger:2007jr,Gustavsson:2007vu}, admits a 3-Lie algebra double copy (with quartic graphs), yielding maximally supersymmetric ${\cal N}=16$ supergravity in 3D. The BLG theory is related to the ${\cal N}=6$ Aharony-Bergman-Jafferis-Maldacena (ABJM) theory~\cite{Aharony:2008ug} for the case of gauge group $SO(4) \sim SU(2) \times SU(2)$.  Both BLG and ABJM theories give partial amplitudes that obey the exotic 3-Lie algebra color-kinematics duality up to six points, and starting at eight points only the BLG theory gives amplitudes that double copy to ${\cal N}=16$ supergravity~\cite{Huang:2012wr}. Much more recently, there has been renewed interest in the double copy applied to 3D massive theories. A new massive double copy has been applied to topologically massive 3D Yang-Mills theory~\cite{Moynihan:2020ejh,Burger:2021wss,Emond:2021lfy,Gonzalez:2021bes,Moynihan:2021rwh,Hang:2021oso,Gonzalez:2021ztm}, which includes a Chern-Simons term in the Lagrangian. In the current paper, we will not directly make use of these quite different constructions, but instead follow the standard formulation~\cite{Bern:2008qj}. 

It is generally accepted that realizing $\mathcal{N}\ge 4$ supersymmetry in Chern-Simons theories requires that the matter transform in a bi-fundamental representation. In our approach, we construct purely-adjoint Chern-Simons-matter theories from imposing that their partial amplitudes obey the BCJ amplitude relations. By including the maximal number of scalars and fermions, we land on a unique theory that exhibits $\mathcal{N}=4$ supersymmetry, at the minor cost of having anticommuting scalars and commuting fermions. While this could potentially be tied to some issue with these theories, it is of little consequence to our construction, since the double copy of two such states produce matter with correct statistics. In particular, we obtain maximally supersymmetric $\mathcal{N}= 8$ Dirac-Born-Infeld theory from the double copy of two  $\mathcal{N}=4$ Chern-Simons-matter theories.

This paper is organized as follows: In section~\ref{Section2}, we introduce Chern-Simons theory and the color-kinematics duality. In sections~\ref{sec:offshellCK} and~\ref{superfield_section},  we discuss off-shell color-kinematics duality for pure Chern-Simons theory, and show that it holds for both the physical fields and the ghost sectors, and we identify an underlying kinematic algebra. In section~\ref{Section5}, we consider off-shell double copies with Chern-Simons theory. In section~\ref{Section6}, we study Chern-Simons-matter theories, and show that they can be made to obey the BCJ relations and that the maximally supersymmetric double copy is an interesting theory.

\section{Preliminaries}
\label{Section2}

We begin with a brief review to introduce the needed background and to set the notation.

\subsection{Chern-Simons theory}

The 3D Chern-Simons action can be written as
\begin{equation} \label{CSaction}
S = \frac{k}{4\pi}\int {\rm Tr} \Big(A\wedge d A + \frac{2i}{3}A\wedge A\wedge A \Big)  \ ,
\end{equation}
where $k$ is the Chern-Simons level, and the gauge field $A$ is a Lie-algebra valued one-form, $A = T^a A_\mu^a dx^\mu$.
Solutions to the corresponding field equation are flat connections, $F=0$,  where $F$ is the field strength two-form, $F = dA + i A\wedge A$.  
Without loss of generality, we take the gauge group to be $SU(N_c)$, with generators and structure constants normalized such that $\textrm{Tr}({T}^a {T}^b)=\frac{1}{2}\delta^{ab}$ and $[{T}^a,{T}^b]=i{f}^{abc}T^c$.

The Chern-Simons action is invariant under infinitesimal gauge transformations $\delta A_\mu = D_\mu \alpha = \partial_\mu \alpha +i[A_\mu,\alpha]$ and must be gauge-fixed in order for the kinetic term to have an inverse.  Introducing the Faddeev-Popov ghosts $c$, $\bar{c}$, and the gauge-fixing functional (for the family of $R_\xi$ gauges)
\begin{eqnarray}\label{eqn:cs_fp}
S_{\rm gf} &=& \frac{k}{2\pi} \int d^3x \,{\rm Tr} \Big(
 -\frac{1}{2\xi}\partial \cdot A\partial\cdot A
  -\bar{c} \Box c - i\bar{c}\, \partial_\mu[A^\mu,c] \Big) \,,
\end{eqnarray}
the kinetic term of the combined action $S+S_{\rm gf}$ can now be inverted. It is easiest to work in the $\xi\rightarrow 0$ limit, giving the  Chern-Simons propagator in Lorenz gauge ($\partial \cdot A=0$),
\begin{equation}
G^{\mu\nu}(x) = \int \frac{d^3p}{(2\pi)^3}\,\frac{\epsilon^{\mu\nu\rho}p_\rho}{p^2{+}i0}\,e^{ip\cdot x} \ .
\end{equation}
For the purpose of considering amplitudes, it is appropriate to study linearized solutions satisfying the Lorenz gauge condition. The linearized equation of motion
\begin{equation}
dA= 0 \ ,
\end{equation}
has the solution
\begin{equation} \label{planewave}
A_\mu^a = p_\mu e^{ip\cdot x} C^a \ , \ \ \textrm{with}\ \ \ p^2=0 \ ,
\end{equation}
where $C^a$ is a color wavefunction. 

The solution (\ref{planewave}) is simply a linearized gauge transformation, as expected since pure Chern-Simons theory has no local degrees of freedom. If one were to compute an amplitude with such external on-shell states, the result would vanish since the  amplitude is gauge invariant. In order to study non-vanishing amplitudes, one has to consider matter particles for all the asymptotic states, i.e. Chern-Simons-matter theories.  That said, pure Chern-Simons theory still has non-trivial correlation functions in terms of the off-shell gauge field, and more importantly the Wilson loops of the theory compute topological knot invariants~\cite{Witten:1988hf}. The perturbative Feynman-diagram expansion can be readily used for the Wilson loop calculations~\cite{bar1995vassiliev,birman1993knot} (for a review see \cite{Labastida:1998ud}). We will not consider Wilson loops in this paper, instead we will focus on the amplitudes approach, and the corresponding off-shell correlation functions.

\subsection{Color-kinematics duality}
Amplitudes in a generic gauge theory can be organized as sums over cubic graphs~\cite{Bern:2008qj,Bern:2010ue}, at tree level the $n$-point formula is
\begin{equation} \label{ampl_n}
{\cal A}_n = g^{n-2}\sum_i \frac{n_i c_i}{D_i} \ ,
\end{equation}
where $n_i$ is the kinematic numerator of the $i$'th graph.  The numerators capture the dependence on local kinematic data, such as momenta, polarizations and possibly flavor. For a purely-adjoint and massless theory, the color factors $c_i$  are products of the gauge group structure constants, re-scaled as $\tilde{f}^{abc}=i\sqrt{2}f^{abc}$, and the denominators $D_i$ are products squared internal momenta $\prod p^2$. For Chern-Simons theory the gauge coupling constant is $g=\sqrt{4\pi/k}$.

Due to the Jacobi identity of the gauge-group Lie algebra, one can find triplets of color factors whose sum vanishes
\begin{equation}
\tilde{f}^{abx}\tilde{f}^{xcd}+\textrm{cyclic}(a,b,c) = 0 ~~~ \Rightarrow ~~~ c_i + c_j + c_k = 0 \ .
\end{equation}
The central statement of the color-kinematics duality is that one can re-organize the amplitude~(\ref{ampl_n}) such that for each such color identity, the corresponding triplet of numerators vanishes~\cite{Bern:2008qj,Bern:2010ue},
\begin{equation}
c_i + c_j + c_k = 0 ~~~ \Rightarrow ~~~ n_i + n_j + n_k = 0 \ .
\end{equation}
Armed with such numerators, it is possible to recycle them and obtain new amplitudes for some ``double-copy'' theory, that is automatically invariant under linearized diffeomorphisms~\cite{Chiodaroli:2017ngp}. With two independent sets of such numerators, the tree-level double copy is defined as~\cite{Bern:2008qj,Bern:2010ue},
\begin{equation} \label{doubleCopy}
{\cal M}_n = \left(\frac{\kappa}{2}\right)^{n-2}\sum_i \frac{n_i \tilde n_i}{D_i} \ ,
\end{equation}
where $\kappa$ is the coupling constant in the resulting theory. If there are dynamical spin-1 gauge fields contained within each numerator, then the double copy gives a gravitational theory with a dynamical spin-2 field. Anticipating the Chern-Simons construction, we note that the double copy may not be readily identifiable with a gravitational theory when the on-shell spin-2 states are absent. 

The possible existence of duality-satisfying numerators can be indirectly checked by so-called {\it BCJ amplitude relations} between color-ordered amplitudes~\cite{Bern:2008qj}, given that the external states are on-shell, adjoint and massless.\footnote{Massive states obtained from spontaneous symmetry breaking still admit BCJ relations~\cite{Chiodaroli:2015rdg}, and non-adjoint matter may preserve a subset of the BCJ relations~\cite{Johansson:2015oia}.} Color-ordered (or partial) amplitudes are the kinematic coefficients that multiply the independent color trace-factors in the full amplitude~(see e.g. review~\cite{Dixon:1996wi}),
\begin{equation}
\mathcal{A}_n = g^{n-2}\sum_{\sigma\in S_{n-1}} {\rm Tr}(\tilde{T}^{a_1}\tilde{T}^{a_{\sigma(2)}}\cdots \tilde{T}^{a_{\sigma(n)}})A_n(1,\sigma(2),\ldots, \sigma(n)) \ , 
\end{equation}
where through $\tilde{f}^{abc}= {\rm Tr}([\tilde{T}^a, \tilde{T}^b] \tilde{T}^c)$ the generators $\tilde{T}^a = \sqrt{2}T^a$ appeared. 

The BCJ relations are linear relations between these partial amplitudes, with rational coefficients depending on the Mandelstam variables $s_{ij}=(p_i+p_j)^2$.  For example, the simplest of these relations at four points takes the form
\begin{equation}
A_4(1,2,4,3) = \frac{s_{14}}{s_{24}}A_4(1,2,3,4) \ ,
\end{equation}
and a simple family of $n$-point BCJ relations can be written as~\cite{Bern:2008qj} 
\be \label{BCJrel}
\sum_{i=1}^{n-2} (p_n \cdot p_{1\cdots i}) \, A_n(1,\ldots, i, n, i+1, \ldots, n-1) =0\ ,
\ee
where $p_{1\cdots i}=p_1+\cdots+p_i$. These ``fundamental BCJ'' relations  were shown to generate all other BCJ relations in ref.~\cite{Feng:2010my}. 

Color-ordered amplitudes that satisfy the BCJ relations can be used directly to obtain the double-copy amplitude, as in the original KLT formula~\cite{Kawai:1985xq},
\begin{equation}
\mathcal{M}_n =  \left(\frac{\kappa}{2}\right)^{n-2} \!\!\! \!\!\! \sum_{\sigma ,\rho\in S_{n-3}(2,\ldots ,n-2)} \!\!\! \!\!\! A_n(1,\sigma,n-1,n)S[\rho|\sigma]\widetilde{A}_n (1,\rho,n,n-1) \ ,
\end{equation}
where $S[\rho|\sigma]$ is the KLT kernel (see refs.~\citep{Bern:1998sv,Bjerrum-Bohr:2010diw,Bjerrum-Bohr:2010kyi,Bern:2019prr} for its explicit form). 
As we will show, this double-copy formula can be used for certain Chern-Simons matter theories, whereas the more general formula~(\ref{doubleCopy}) will be used when amplitudes cannot be defined on-shell without vanishing. 
In the remaining sections, we will suppress the explicit appearance of coupling constants.

\section{Off-shell color-kinematics duality}\label{sec:offshellCK}
In this section we introduce and study off-shell color-kinematics duality, and apply it to the gauge-field sector of Chern-Simons theory. In the next section, the Faddeev-Popov ghosts will also be included.  

\subsection{Generic off-shell formulation }

Off-shell color-kinematics duality simply means that there exists a complete set of Feynman rules for the theory that obey the duality without imposing on-shell conditions on external lines. The duality has to hold for all fields in the theory, such that the Feynman diagrams that involve sums over fields will obey the duality by the superposition principle. Hence, by construction, off-shell color-kinematics duality should imply that loop diagrams satisfy the duality.  

Let us start by considering a generic gauge theory that describes the adjoint field $\Phi$, which may be a superfield (superposition of fields). Assume that we can pick some gauge such that the field equations take the form
\begin{equation}
\Box \Phi^a =if^{abc}\, bV(\Phi^b,\Phi^c) \ ,
\end{equation}
where via $\Phi^a = 2{\rm Tr} (\Phi T^a)$ we exposed the adjoint indices. Here $V$ is the kinematic part of the three-point Feynman vertex, written as an antisymmetric bilinear 2-to-1 operator, and $b$ is a linear operator that can be identified as the numerator of the propagator $b\sim \Box G$. 

To check the duality, consider three distinct fields $\Phi_1$, $\Phi_2$, $\Phi_3$, which by the superposition principle can be taken to be off-shell plane waves in the chosen gauge with color dependence stripped off. Then off-shell color-kinematics duality is satisfied given that the nested $bV$-product of fields obeys the kinematic Jacobi identity,
\begin{equation}\label{abstractjacobi}
bV(b V(\Phi_1, \Phi_2),\Phi_3) + {\rm cyclic}(1,2,3) = 0 \ ,
\end{equation}
which is a quadratic constraint on the $bV$ operator. 

Now consider the computation of kinematic numerators at multiplicity $n$. Assuming we have computed nested $bV$-products of the fields $\Phi_1,\ldots,\Phi_{n-1}$, one further needs to compute the overlap with a final $\Phi_n$ as external field, $\langle V(\ldots), \Phi_n\rangle$, where the last $b$ is amputated. The overlap is an integration over coordinates, accompanied with possible Lorentz and flavor contractions
$\langle\Phi,\Phi' \rangle=\int d^Dx \, \Phi \cdot\Phi' $. 
For example, a four-point diagram numerator would be given by
\begin{equation}
\begin{tikzpicture}[baseline={(0, -0.1cm)}]
\draw[thick] (-0.33,0) -- (0.33,0);
\draw[thick] (-0.5,0.5) -- (-0.33,0);
\draw[thick] (-0.5,-0.5) -- (-0.33,0);
\draw[thick] (0.5,0.5) -- (0.33,0);
\draw[thick] (0.5,-0.5) -- (0.33,0);
\node at (-0.5*1.25,-0.5*1.25) {$1$};
\node at (-0.5*1.25,0.5*1.25) {$2$};
\node at (0.5*1.25,0.5*1.25) {$3$};
\node at (0.5*1.25,-0.5*1.25) {$4$};
\end{tikzpicture} 
=\,
\langle V( bV(\Phi_1, \Phi_2),\Phi_3) , \Phi_4 \rangle  \ ,
\end{equation}
and the five-point numerator is
\begin{eqnarray}
\begin{tikzpicture}[baseline={(0, 0.2cm)}]
\draw[thick] (-1,0) -- (1,0);
\draw[thick] (0,0) -- (0,0.5);
\draw[thick] (0.5,0) -- (0.5,0.5);
\draw[thick] (-0.5,0) -- (-0.5,0.5);
\node at (-1.25,0) {$1$};
\node at (-0.5,0.7) {$2$};
\node at (0,0.7) {$3$};
\node at (0.5,0.7) {$4$};
\node at (1.25,0) {$5$};
\end{tikzpicture} 
\, &=&\,
\langle V(bV( bV(\Phi_1, \Phi_2),\Phi_3) , \Phi_4 ),\Phi_5 \rangle  \ .
\end{eqnarray}

As a word of caution, one might encounter a situation where \eqn{abstractjacobi} holds, but the amputated expression $V( bV(\Phi_1, \Phi_2),\Phi_3) + {\rm cyclic}(1,2,3)$ does not vanish, instead belonging to the kernel of the operator $b$. We will see that for the case that we are interested in, pure Chern-Simons theory, the vanishing of the amputated object can be enforced by making an appropriate gauge choice.

\subsection{Tree level Chern-Simons correlators}

From the general considerations in the previous subsection, we now apply the formalism to pure Chern-Simons theory. Starting from the field equation, $F=0$, written out as 
\begin{equation}
\epsilon^{\mu\nu\rho} \partial_\nu A^a_\rho = -i f^{abc}\epsilon^{\mu\nu\rho}A^b_\nu A^c_\rho \ ,
\end{equation}
we can contract both sides of the equation with the propagator numerator $b_{\sigma\rho}\equiv\epsilon_{\sigma\rho\alpha}\partial^\alpha$. Assuming that the vector field is in the Lorenz gauge $\partial\cdot A= 0$, we obtain
\begin{equation}
\square A_\sigma = i b_{\sigma\mu}\epsilon^{\mu\nu\rho} f^{abc}A^b_\nu A^c_\rho \ ,
\end{equation}
meaning that the vertex function is
\begin{equation}
V(A_{1},A_{2}) = \epsilon^{\rho A_1 A_2} \ ,
\end{equation}
where the Schoonship notation $\epsilon^{\rho A_1 A_2}\equiv \epsilon^{\rho \mu\nu}A_{1\mu}A_{2\nu}$ is used for simplicity.

In the standard formulation of tree-level Chern-Simons theory there is a single field, the gauge field $A$, the requirement to have off-shell color-kinematics duality is then that the $b$ operator and the vertex $V$ obey a Jacobi identity.
This will imply that the theory obeys the duality for any tree level diagram (for loops, one must consider the ghost sectors too, which are treated in the next section). For this purpose we study the off-shell four-point correlation function. We note that the external fields in the correlation function are sourced by some current through a propagator\footnote{Unlike the case of amplitudes, the external propagators in the correlation functions are not amputated. }, which will automatically enforce Lorenz gauge
\begin{equation}
A^\mu = \frac{b^{\mu \nu}}{\Box}J_\nu 
~~~\Rightarrow ~~~
\partial_\mu A^\mu = 0 \ .
\end{equation}
It is useful to make the same statement in momentum space, except we only consider the numerator of the propagator 
\begin{equation}
\varepsilon^\mu(p) \equiv \epsilon^{\mu\nu\rho} p_{\nu}\epsilon_\rho(p) 
~~~\Rightarrow ~~~
p_{\mu} \varepsilon^\mu(p)= 0 \ ,
\end{equation}
where $\epsilon^\rho(p)$ is the Fourier transform of the current, and $\varepsilon^\mu(p)$ is the Fourier transform of $\Box A^\mu$. Note that the current is not necessarily conserved off shell, and hence $\epsilon^\rho(p)$ is an unconstrained function. 

All together, the $s$-channel numerator for the off-shell four-point correlation function takes the form 
\begin{equation}
\begin{tikzpicture}[baseline={(0, -0.1cm)}]
\draw[thick] (-0.33,0) -- (0.33,0);
\draw[thick] (-0.5,0.5) -- (-0.33,0);
\draw[thick] (-0.5,-0.5) -- (-0.33,0);
\draw[thick] (0.5,0.5) -- (0.33,0);
\draw[thick] (0.5,-0.5) -- (0.33,0);
\node at (-0.5*1.25,-0.5*1.25) {$1$};
\node at (-0.5*1.25,0.5*1.25) {$2$};
\node at (0.5*1.25,0.5*1.25) {$3$};
\node at (0.5*1.25,-0.5*1.25) {$4$};
\end{tikzpicture} 
=
 \epsilon^{\varepsilon_1 \varepsilon_2 \nu}\epsilon_{\nu p_{12} \mu}\epsilon^{\mu \varepsilon_3 \varepsilon_4} \ .
\end{equation}
In order to prove that this object obeys the color-kinematics duality, it is convenient to re-express it using the identity $ \epsilon^{\mu_1\mu_2\mu_3} \epsilon_{\nu_1\nu_2\nu_3}= 3!\, \delta^{[\mu_1}_{\nu_1} \delta^{\mu_2}_{\nu_2} \delta^{\mu_3]}_{\nu_3} $, as well as momentum conservation and transversality $\varepsilon_i\cdot p_i=0$, giving
\begin{equation}
2 \epsilon^{\varepsilon_4 \varepsilon_3 \nu}\epsilon_{\nu p_{12} \mu}\epsilon^{\mu \varepsilon_2  \varepsilon_1} =
\varepsilon_4\cdot p_{3}\epsilon^{ \varepsilon_1  \varepsilon_2 \varepsilon_3} 
-\varepsilon_{1}\cdot p_{2}\epsilon^{\varepsilon_2  \varepsilon_3 \varepsilon_4}
-\varepsilon_2\cdot p_{1}\epsilon^{\varepsilon_3\varepsilon_1\varepsilon_4}
-\varepsilon_3\cdot p_{4} \epsilon^{ \varepsilon_1\varepsilon_2 \varepsilon_4}
 \ .
\end{equation}
In this form it is not hard to see that the cyclic sum $cyclic(1,2,3)$ vanishes, and hence the off-shell BCJ numerator obeys the Jacobi identity. Since we made no assumptions about the momentum or the external sources, the above $s$-channel numerator can be inserted into any multiparticle tree-level (or loop-level) diagram and the Jacobi identity will hold.  Along similar lines, it can be shown that correlation functions involving Faddeev-Popov ghosts also satisfy the Jacobi identity, but this will naturally emerge once we employ the superfield notation in section~\ref{superfield_section}.

\subsection{Generators for the kinematic algebra}\label{sec:first_KA}
Having established that the standard formulation of Chern-Simons theory (without Faddeev-Popov ghosts) obeys the color-kinematics duality, it should now be possible to find an explicit representation of the  kinematic algebra. Indeed, we can define momentum-dependent generators 
\begin{equation} \label{vectorGen}
L^\mu(p) = e^{ip\cdot x}\Delta^{\mu\nu} \partial_\nu \ ,
\end{equation}
where $\Delta^{\mu\nu}$ is a transversality-enforcing projector, identified with the Fourier-transform of the $b$ operator, 
\begin{equation}
\Delta^{\mu\nu}(p)=i\epsilon^{\rho \mu\nu} p_\rho \ .
\end{equation}
The generators form an infinite-dimensional Lie algebra,
\begin{eqnarray} \label{kinalg}
[L^\mu(p_1),L^\nu(p_2)] &=& F^{\mu\nu}_{\ \ \rho}\, L^\rho(p_1+p_2)   \ ,
\end{eqnarray}
where the $F^{\mu \nu}_{\ \ \rho}$ are the structure constants, which also depend on the two momenta corresponding to the upper indices,
\begin{equation}
F^{\mu_1\mu_2}_{\phantom{\mu_1\mu_2} \nu}(p_1,p_2) = \Delta^{\rho\mu_1}(p_1) \epsilon_{\rho\nu\sigma} \Delta^{\sigma\mu_2}(p_2) \ .
\end{equation}
One can easily verify that the structure constants are related to the Chern-Simons three-point interaction, contracted with two propagator numerators in momentum space.  

To obtain BCJ numerators  from this kinematic algebra, we simply compute nested commutators of the generators. For example, the numerator corresponding to five off-shell Chern-Simons fields (in momentum space) is
\begin{eqnarray}
\begin{tikzpicture}[baseline={(0, 0.2cm)}]
\draw[thick] (-1,0) -- (1,0);
\draw[thick] (0,0) -- (0,0.5);
\draw[thick] (0.5,0) -- (0.5,0.5);
\draw[thick] (-0.5,0) -- (-0.5,0.5);
\node at (-1.25,0) {$1$};
\node at (-0.5,0.7) {$2$};
\node at (0,0.7) {$3$};
\node at (0.5,0.7) {$4$};
\node at (1.25,0) {$5$};
\end{tikzpicture} 
\, &=&\,
{\rm tr} \Big([[[L^{\mu_1}(p_1),L^{\mu_2}(p_2)],L^{\mu_3}(p_3)],L^{\mu_4}(p_4)], L^{\mu_5}_{\rm amp}(p_5)\Big)  \\
&=&
F^{\mu_1\mu_2}_{\hspace{0.7cm}\nu}F^{\nu \mu_3}_{\hspace{0.5cm}\rho}F^{\rho \mu_4 \mu_5} \delta^{3}(p_1{+}p_2{+}p_3{+}p_4{+}p_5) \ ,
\end{eqnarray}
where, in order to avoid over-counting projectors, the last generator  $L^\mu_{\rm amp}(p) = e^{ip\cdot x} \partial^\mu$ is amputated. We take the ${\rm tr}(\cdots)$ operation to be formally defined as
\be
{\rm tr}\big(L^{\mu}(p)  L^{\nu}_{\rm amp}(q)\big) \equiv \delta^{3}(p+q)\Delta^{\mu \nu}\,,
\ee
and the structure constants with three raised indices are then $F^{\rho \mu_4 \mu_5} = F^{\rho \mu_4}_{\phantom{\rho \mu_4}\nu} \Delta^{\nu \mu_5}(p_5)$. As is apparent, the above kinematic numerator includes  the momentum conserving delta function, but by convention we strip off this factor before inserting the numerator into the amplitude or correlation function.

The kinematic Lie algebra (\ref{kinalg}) generates volume-preserving diffeomorphisms in three dimensions. Consider two unconstrained functions: a scalar function $f(x)$ and vector function $\epsilon_\mu(p)$. Then the action of a finite group element is
\be
 e^{\int d^3p \, \epsilon_\mu(p) L^\mu(p)  } f(x)  =  f(X(x))  \,,
\ee
where the operator in the exponent evaluates to
\be
 \int d^3p \,\epsilon_\mu(p) L^\mu(p) = \tilde A_\mu(x)  \partial^\mu\,,
\ee
which is a diffeomorphism generator. And the finite shift of the argument of the funtion $f(x)$ can be worked out from $X^\mu(x) = e ^{\tilde A \cdot  \partial} x^\mu = x^\mu+ \tilde A^\mu(x) + {\cal O}(\tilde A^2)$.
By construction $\tilde A_\mu(x)$ is a generic divergence-free vector field, $\partial \cdot\tilde A=0$, and it follows that the determinant of the Jacobian is preserved for any such transformation,  $\det(\partial_\mu X^\nu) =1$, hence the algebra is that of volume preserving diffeomorphisms. Note that the Fourier transform of the vector is the transverse (off-shell) polarization vector that we used before $\varepsilon_\mu(p) \equiv \Delta_{\mu \nu}(p) \epsilon^\nu(p)$. 

We will discuss the diffeomorphism algebra in more detail in section~\ref{sec:ck}, where the kinematic algebra is extended to also include the Faddeev-Popov ghosts.

\section{Chern-Simons in superfield notation}\label{superfield_section}
In order to streamline the loop-level discussion of Chern-Simons color-kinematics duality, we consider the superfield formulation that collects together the ghosts and vector fields. 

\subsection{The superfield notation}
The Faddeev-Popov action for Chern-Simons theory admits a convenient superfield notation as shown by Axelrod and Singer~\cite{Axelrod:1991vq}. Consider the BRST action including the Nakanishi-Lautrup gauge fixing field $B$,
\begin{equation} 
S_{\rm gf} = \frac{k}{2\pi} \int d^3x \, \textrm{Tr} \left(
 \frac{\xi}{2}B^2
  -B \partial\cdot A -\bar{c} \, \Box c - i\bar{c}\, \partial_\mu[A^\mu,c] \right) \, .
\label{eqn:chern-simons-action}
\end{equation} 
Integrating out the $B$ field gives the gauge-fixing term in equation \eqref{eqn:cs_fp}. Alternatively, one can first take the $\xi\rightarrow 0$ limit in \eqref{eqn:chern-simons-action}, which gives
\begin{equation} 
S_{\rm gf} = \frac{k}{2\pi} \int d^3x \, \textrm{Tr} \left( -B \, \partial\cdot A -\bar{c} \, \Box c - i\bar{c}\,\partial_\mu[A^\mu,c] \right) \, .
\end{equation} 
Since the $B$ field appears linearly one can integrate it out to obtain the (metric dependent) Lorenz gauge $\partial\cdot A=0$. Further, one notes that the $\bar{c}$ field always appears with its derivative, suggesting the introduction of the dualized field $C^{\mu\nu}=\frac{1}{2}\epsilon^{\mu\nu\rho}\partial_\rho \bar{c}$, giving the full Chern-Simons action 
\begin{equation} \label{fullL}
S = \frac{k}{2\pi}\int {\rm Tr}\left(
\frac{1}{2}A\wedge d A - C\wedge d c + \frac{i}{6}(A\wedge A\wedge A - 6 \, C\wedge [A,c]) 
\right)\, .
\end{equation}
The action is written in terms of forms and exterior derivatives and it is possible to introduce Grassmann-odd variables $\theta^\mu$ that allow the forms to combine into a single superfield, 
\begin{equation}
\Psi = c + \theta_\mu A^\mu + \theta_\mu \theta_\nu {C}^{\mu\nu} + \theta_1\theta_2\theta_3 a \ .
\end{equation}
The last field $a$ was not present in the action (\ref{fullL}), and here it is introduced for completeness of the discussion. It corresponds to a non-dynamical field, and for our purpose we can assume $a=0$, which is compatible with Lorenz gauge. 

Using the superfield notation, the Chern-Simons action becomes
\begin{equation} \label{superCS}
S = \frac{k}{2\pi}\int d^3x d^3\theta \, {\rm Tr}\Big(
 \frac{1}{2}\Psi Q\Psi + \frac{i}{3}\Psi\Psi\Psi 
\Big)
 \, ,
\end{equation}
where $Q=\theta_\mu\partial^\mu$ has the interpretation of a super-differential, or alternatively, a BRST operator for the worldline action\footnote{The superfield formulation of Chern-Simons theory is reminiscent of the more recent formulation of 10D super-Yang-Mills in terms of a pure-spinor superparticle~\citep{Berkovits:2001rb}.} of Chern-Simons theory \cite{Berkovits:2001rb}. Forgetting about the steps in the above derivation, the new action (\ref{superCS}) can be taken as defining the quantum Chern-Simons theory, where all fields are now unconstrained. As such, the action respects local gauge invariance, which in the superfield notation corresponds to the infinitesimal transformation 
\begin{equation}
\delta \Psi = Q\Omega + i[\Psi,\Omega] \ ,
\end{equation}
for some superfield $\Omega$.

In perturbation theory the action needs to be gauge fixed. The gauge condition and Feynman rules in the superfield notation make use of the linear operator $b$,\footnote{The notation is inspired by the pure-spinor Feynman rules used in ref.~\cite{Ben-Shahar:2021doh}, where the corresponding operator is the composite worldline $b$ ghost.}
 now defined as
\begin{equation}
b = \frac{\partial}{\partial\theta^\mu}\partial_\mu \equiv \bop^\mu \partial_\mu \ ,
\end{equation}
which is the codifferential in superspace. It anti-commutes with $Q$ into the d'Alembertian, 
\begin{equation}\label{hodge}
bQ+Qb = \square \ ,
\end{equation} 
which is known as Hodge decomposition. It is nilpotent $b^2=0$ and, since it contains two derivatives, it obeys the second-order Leibniz rule. 

The Lorenz gauge condition for a super-field now becomes
\begin{equation}
b\Psi = 0 \ . 
\end{equation}
This imposes no constraint on the $c$ ghost, and Lorenz gauge on the vector $\partial_\mu A^\mu=0$ and on the 2-form $\partial_\mu C^{\mu\nu}= 0$. The last constraint implies that the 2-form ghost can always be solved as $C^{\mu\nu}=\frac{1}{2}\epsilon^{\mu\nu\rho}\partial_\rho \bar{c}$. Furthermore, Lorenz gauge implies that the $a$ field is constant, and we set it to zero $a=0$. 

\subsection{Superspace Feynman rules}
To establish that color-kinematics duality holds off shell in pure Chern-Simons theory, including for all loop diagrams, we now consider the needed Feynman rules. The superfield Feynman rules, which automatically include ghosts in the loops, are  
\begin{eqnarray} \label{FeynmanRules}
\begin{tikzpicture}[baseline={(0, -0.1cm)}]
\draw[thick] (0,0) -- (1,0);
\node at (-0.25,0) {$\theta$};
\node at (1+0.25,0) {$\tilde \theta$};
\node at (0.5,0.1) {$\to$};
\node at (0.5,-0.25) {$p$};
\end{tikzpicture}
=
\frac{p \cdot \bop}{p^2}\delta^3(\theta - \tilde \theta) \ ,
\hspace{2cm}
\begin{tikzpicture}[baseline={(0, -0.1cm)}]
\draw[thick] (0,0.75) -- (0,0);
\draw[thick] (0,0) -- (0.87*0.75,-0.5*0.75);
\draw[thick] (0,0) -- (-0.87*0.75,-0.5*0.75);
\node at (0.05,-0.3) {$\theta$};
\end{tikzpicture}
~~=~~
i\int d^3\theta \ ,
\end{eqnarray}
where the arrow indicates the momentum flow, and the internal vertices are labeled by distinct $\theta$s. We remind the reader that $\bop^\mu = \frac{\partial}{\partial \theta_\mu}$ is the conjugate to $\theta^\mu$, and that we have suppressed the coupling constant and momentum conserving delta functions. The Feynman rules are in a hybrid form:  the $x$'s are Fourier transformed but not the $\theta$ coordinates. 

Next, we consider the numerator of an off-shell four-point diagram, it is
\begin{equation} \label{4ptExample}
\begin{tikzpicture}[baseline={(0, -0.1cm)}]
\draw[thick] (-0.33,0) -- (0.33,0);
\draw[thick] (-0.5,0.5) -- (-0.33,0);
\draw[thick] (-0.5,-0.5) -- (-0.33,0);
\draw[thick] (0.5,0.5) -- (0.33,0);
\draw[thick] (0.5,-0.5) -- (0.33,0);
\node at (-0.5*1.25,-0.5*1.25) {$1$};
\node at (-0.5*1.25,0.5*1.25) {$2$};
\node at (0.5*1.25,0.5*1.25) {$3$};
\node at (0.5*1.25,-0.5*1.25) {$4$};
\end{tikzpicture} 
=
\int d^3 \theta \Psi_1 \Psi_2 \int d^3 \tilde{\theta} \, p_{34} \cdot \bop \, \delta^3(\theta-\tilde\theta)\Psi_3( \tilde{\theta})\Psi_4( \tilde{\theta})  \ ,
\end{equation}
where the external wavefunctions are assumed to satisfy the gauge condition $b\Psi=0$.  Given that the external fields are sourced at some external point and then propagate to the interaction vertex, $\Psi =\frac{b}{\Box} {\cal J}$, Lorenz gauge is automatic by nilpotency of the $b$ operator.
 
After evaluating the $\tilde\theta$ integral, and using integration by parts and momentum conservation, the above diagram simplifies to 
\begin{equation}
\begin{tikzpicture}[baseline={(0, -0.1cm)}]
\draw[thick] (-0.33,0) -- (0.33,0);
\draw[thick] (-0.5,0.5) -- (-0.33,0);
\draw[thick] (-0.5,-0.5) -- (-0.33,0);
\draw[thick] (0.5,0.5) -- (0.33,0);
\draw[thick] (0.5,-0.5) -- (0.33,0);
\node at (-0.5*1.25,-0.5*1.25) {$1$};
\node at (-0.5*1.25,0.5*1.25) {$2$};
\node at (0.5*1.25,0.5*1.25) {$3$};
\node at (0.5*1.25,-0.5*1.25) {$4$};
\end{tikzpicture} 
= i \int d^3 \theta b(\Psi_1 \Psi_2) \Psi_3\Psi_4 \ .
\end{equation}
Here the $b$ operator can be represented in momentum or position space, $b=\bop \cdot  \partial= ip \cdot \bop$, depending on what is convenient, and it only acts on the fields inside the parenthesis.

In general, every tree-level diagram can be reduced to a single integral over superspace coordinates acting on a function made of nested $b$ operators and fields. For example, the five-point half ladder diagram can be written as
\begin{equation}
\begin{tikzpicture}[baseline={(0, 0.2cm)}]
\draw[thick] (-1,0) -- (1,0);
\draw[thick] (0,0) -- (0,0.5);
\draw[thick] (0.5,0) -- (0.5,0.5);
\draw[thick] (-0.5,0) -- (-0.5,0.5);
\node at (-1.25,0) {$1$};
\node at (-0.5,0.7) {$2$};
\node at (0,0.7) {$3$};
\node at (0.5,0.7) {$4$};
\node at (1.25,0) {$5$};
\end{tikzpicture}  =i \int d^3 \theta \, b(b(\Psi_1 \Psi_2)\Psi_3)\Psi_4\Psi_5\ .
\end{equation} 
The combination of the superspace integration and the derivatives inside the propagator-numerator, $b$, correctly enforce all the Lorentz-index and field contractions making the calculation equivalent to using the Feynman rules (\ref{FeynmanRules}). Note that the number of $b$ insertions precisely matches the number of internal lines. 

Loop diagrams are most straightforwardly calculated by following the Feynman rules directly, with integrations over $\theta$s for each internal vertex, as done in \eqn{4ptExample}. However, they can also be obtained by recycling the above simple formulas for the tree numerators, and gluing together pairs of external off-shell legs. Consider the one-loop box diagram, split into a six-point tree graph with an extra pair of fields with momenta $\pm\ell$,
\begin{equation}
\begin{tikzpicture}[baseline={(0, 0.4cm)}]
\draw[thick] (-1,0) -- (1,0);
\draw[thick] (-1,1) -- (1,1);
\draw[thick] (0.5,0) -- (0.5,1);
\draw[thick] (-0.5,0) -- (-0.5,1);
\node at (-1.25,0) {$1$};
\node at (-1.25,1) {$2$};
\node at (0,0.25) {$\ell$};
\node at (1.25,0) {$4$};
\node at (1.25,1) {$3$};
\end{tikzpicture}  ~= ~
\begin{tikzpicture}[baseline={(0, 0.2cm)}]
\draw[thick] (-1,0) -- (1.5,0);
\draw[thick] (0,0) -- (0,0.5);
\draw[thick] (0.5,0) -- (0.5,0.5);
\draw[thick] (1,0) -- (1,0.5);
\draw[thick] (-0.5,0) -- (-0.5,0.5);
\node at (-1.35,0) {$-\ell$};
\node at (-0.5,0.7) {$1$};
\node at (0,0.7) {$2$};
\node at (0.5,0.7) {$3$};
\node at (1,0.7) {$4$};
\node at (1.75,0) {$\ell$};
\end{tikzpicture} 
 ~= ~ \int d^3 \theta \, b(b(b(b(\widetilde \Psi_{-\ell})\Psi_1)\Psi_2)\Psi_3)\Psi_4\Psi_{\ell}\ ,
\end{equation}
where we now inserted four $b$ corresponding to the four internal loop lines.  How do we glue together the fields appearing in $\widetilde \Psi_{-\ell}$ and $\Psi_{\ell}$ ? We cannot do it with another $\theta$ integral, because we already integrated out these variables. It turns out that we can replace the fields in $\Psi_{-\ell}$ by functional derivatives, which will act on the $\Psi_{\ell}$ fields to achieve the needed contractions. Hence it is convenient treat $b(\Psi_{-\ell})$ as a special operator field\footnote{An alternative is to introduce creation and annihilation operators for the fields carrying loop momenta, but this goes beyond our needs for  performing simple field contractions.} defined to be
\be
b(\widetilde \Psi_{-\ell})\equiv  \frac{i}{2}\ell_\mu \epsilon^{\mu \nu \rho} \frac{\delta}{\delta  C_\ell^{\nu \rho} }
 -i \ell_\mu \theta_\nu   \epsilon^{\mu \nu \rho} \frac{\delta}{\delta A^\rho_\ell }+ \frac{i}{2} \ell_\mu \theta_\nu \theta_\rho   \epsilon^{\mu \nu \rho} \ \frac{\delta}{\delta c_\ell } \ .
\ee

Higher-loop diagrams can be handled in the same way, and this confirms that loop-level numerators can always be obtained by starting from tree numerators, and then appropriately gluing together pairs of external states. While this procedure appears to not manifestly respect the automorphism symmetries of the loop diagrams (such as the cyclic symmetry of the box), the symmetries are manifest in the Feynman rule calculations, which give the same numerators.   

\subsection{Color-kinematics duality and the superspace kinematic algebra}\label{sec:ck}
To prove that Chern-Simons theory obeys off-shell color-kinematics duality, we repeat the analysis of section~\ref{sec:offshellCK}, but in the superfield notation. Since this includes the Faddeev-Poppov ghosts, the result extends automatically to loop level. 

We here show that color-kinematics duality in Chern-Simons theory follows from the fact that the propagator numerator is a second-order differential,\footnote{See related considerations for Yang-Mills theory in refs.~\cite{Reiterer:2019dys,Ben-Shahar:2021doh}. Specifically, in ref.~\cite{Reiterer:2019dys} for the first-order formulation of Yang-Mills theory,  the propagator-numerator was a perturbed version of the codifferential, called the $h$ operator.}
being a product of two derivatives $b= \bop \cdot \partial$.  The second-order nature of the operator gives the identity
\begin{eqnarray*}
b(\Psi_1\Psi_2\Psi_3) 
&=&b(\Psi_{1}\Psi_2)\Psi_3 - \Psi_1 b(\Psi_{2}\Psi_3) - b(\Psi_{1}\Psi_3)\Psi_2 \\
&&\null
-b\Psi_1\Psi_2\Psi_3 + \Psi_1 b\Psi_2\Psi_3 - \Psi_1\Psi_2 b\Psi_3 \ ,
\end{eqnarray*}
which in Lorenz gauge, $b(\Psi_i) = 0$, simplifies to
\begin{eqnarray} 
b(\Psi_1\Psi_2\Psi_3) 
&=&b(\Psi_{1}\Psi_2)\Psi_3 + b(\Psi_{2}\Psi_3)\Psi_1 +b(\Psi_{3}\Psi_1)\Psi_2 \ .
\end{eqnarray}
Recall that all wavefunctions satisfy the Lorenz gauge condition, since they are assumed to be obtained from propagators (external or internal) that enforce the Lorenz gauge condition.
After multiplying by a fourth field and integrating over $\theta$, the identity becomes
\begin{equation}\label{jacobiEqn}
\int d^3 \theta b(\Psi_1 \Psi_2)\Psi_3\Psi_4 + \textrm{cyclic}(1,2,3) = \int d^3 \theta b(\Psi_1 \Psi_2 \Psi_3)\Psi_4 =0 \,.
\end{equation}
The last equality follows via integration by parts  $ b(\Psi_1 \Psi_2 \Psi_3)\Psi_4 \sim \Psi_1 \Psi_2 \Psi_3b(\Psi_4)$, and it vanishes because of the Lorenz gauge.  
The left hand side of \eqn{jacobiEqn} is precisely the Jacobi identity for a triplet of kinematic numerators. Hence the Chern-Simons Feynman rules obey color-kinematics duality.  

Let us emphasize the result, by considering the Feynman rules directly at loop level.  Assume we have a multi-loop Feynman diagram, and we isolate an internal propagator line that we plan to perform a Jacob identity on. The diagram looks as follows:
\begin{eqnarray}
\begin{tikzpicture}[baseline={(0, -0.1cm)}]
\draw[fill=gray!40] (0,0) ellipse (1.7cm and 1.2cm);
\draw[fill=white] (0,0) ellipse (1cm and 0.6cm);
\draw[thick] (-0.33,0) -- (0.33,0);
\draw[thick] (-0.5,0.5) -- (-0.33,0);
\draw[thick] (-0.5,-0.5) -- (-0.33,0);
\draw[thick] (0.5,0.5) -- (0.33,0);
\draw[thick] (0.5,-0.5) -- (0.33,0);
\node at (-0.5*1.25,-0.5*1.25) {$1$};
\node at (-0.5*1.25,0.5*1.25) {$2$};
\node at (0.5*1.25,0.5*1.25) {$3$};
\node at (0.5*1.25,-0.5*1.25) {$4$};
\end{tikzpicture} 
=
\int d^3 \theta \Psi_1 \Psi_2 \left(\int d^3 \tilde \theta \, p_{34} \cdot \bop \, \delta(\theta-\tilde \theta)\Psi_3(\tilde \theta)\Psi_4(\tilde \theta)\right) \ ,
\end{eqnarray}
where the displayed wavefunctions are sourced by the rest of the diagram, here schematically indicated by the grey blob. The wavefunctions are in Lorenz gauge by virtue of the propagators on lines 1--4 enforcing this. Then since this expression evaluates to the first term in~\eqn{jacobiEqn}, it follows that after adding the remaining two diagrams in the cyclic orbit $(1,2,3)$, we obtain the desired kinematic Jacobi identity.

The obtained Jacobi identity can be attributed to a Poisson algebra which emerges from the action of $b$ on functions in Lorenz gauge. 
The operation of $b$ on two such fields can be written as 
\begin{equation}\label{poisson-bracket}
    b(\Psi_1\Psi_2)~\stackrel{\rm Lorenz}{=}~\bop^\mu\Psi_1\partial_\mu\Psi_2 - \partial_\mu\Psi_1 \bop^\mu\Psi_2 \equiv \{\Psi_1,\Psi_2\}_{\rm P}\ .
\end{equation}
The Jacobi identity is trivialized by this rewriting,
\begin{equation}\label{eqn:poisson_jacobi}
    b(b(\Psi_1\Psi_2)\Psi_3) + \textrm{cyclic}(1,2,3) =\{\{\Psi_1,\Psi_2\}_{\rm P},\Psi_3\}_{\rm P}+ \textrm{cyclic}(1,2,3)= 0 \ .
\end{equation}
From the Poisson bracket it is a small step to define generators labeled by a superfield $\Psi$,
\begin{equation}
    L_\Psi \equiv \bop^\mu\Psi\partial_\mu - \partial_\mu\Psi \bop^\mu \  ,
\end{equation}
which satisfy the Lie algebra
\begin{equation}
[L_{\Psi_1},L_{\Psi_2}] = L_{b(\Psi_1\Psi_2)} \ .
\end{equation}
Similar to the vectorial generators in \eqn{vectorGen}, this is the Lie algebra of a subgroup of 3D superspace diffeomorphisms, with generators restricted to Lorenz gauge $b\Psi=0$. 
The subgroup corresponds to the diffeomorphisms that preserve both the superspace volume form, $d^3x d^3\theta$, and separately the bosonic and fermionic volume forms. This can be shown by considering the action of an infinitesimal diffeomorphism.
Note, while the above Poisson bracket and dual Lie algebra can be defined for any functions, not necessarily in Lorenz gauge, the matching to the superspace Feynman rules of Chern-Simons theory only holds for superfields in the Lorenz gauge.

To make contact with the kinematic algebra discussed in section~\ref{sec:first_KA}, we can take the function that labels the generator to be restricted to a vectorial planewave $\Psi =  e^{ip\cdot x}\varepsilon_\nu(p) \theta^\nu$, then the relation between the two types of generators are
\begin{equation} 
    \epsilon_\mu L^\mu(p) = L_\Psi\Big|_{\bop\rightarrow0}\,,
\end{equation}
where the transverse polarization $\varepsilon_\nu(p) = \Delta_{\nu\mu}\epsilon^\mu$ is given in terms of an unconstrained $\epsilon^\mu$. We also set the Grassmann derivative $\bop$ to zero since the $L^\mu(p)$ generator only acts on Grassmann-even functions.

\section{Double copies with Chern-Simons theory}
\label{Section5}
Since all amplitudes of pure Chern-Simons theory vanish, the corresponding double-copy amplitudes that involve Chern-Simons kinematic numerators will also vanish on shell. Nevertheless, it is still instructive to carry out the double copy off shell, and thus attempt to reconstruct the off-shell actions of the double-copy theories. The resulting Lagrangians have the somewhat disagreeable feature that they are already in a gauge-fixed form, and kinetic terms easily become non local. Thus, in general, further work is needed to better understand the underlying local and gauge invariant Lagrangians.

\subsection{Chern-Simons $\otimes$ Chern-Simons}
The first double copy we consider is Chern-Simons theory with itself, in the standard formulation where we only have the gauge field $A^\mu$. Recall that the propagator for the Chern-Simons field is
\begin{equation}
G^{\mu\nu}(x) = \int \frac{d^3p}{(2\pi)^3} \frac{\epsilon^{\mu\alpha\nu}p_\alpha}{p^2}e^{ip\cdot x} \ .
\end{equation}
The propagator for the field\footnote{The operation $A^\mu \otimes A^{\dot \mu}$ should be thought of as a tensor product of the fields in momentum space~\cite{Bern:2019prr}, or as the convolution of two fields in position space~\cite{Anastasiou:2014qba,Anastasiou:2018rdx,Luna:2020adi}. } $H^{\mu\dot\mu}\equiv  A^\mu \otimes A^{\dot \mu}$ in the double-copy theory is obtained by writing down two copies of the same numerator,
\begin{equation}
G^{\mu\dot{\mu},\nu\dot{\nu}}(x) = \int \frac{d^3p}{(2\pi)^3} \frac{
\epsilon^{\mu\alpha\nu}p_\alpha 
\epsilon^{\dot{\mu}\dot{\alpha}\dot{\nu}}p_{\dot{\alpha}}
}{p^2}e^{ip\cdot x} \ ,
\end{equation}
where the dotted indies are also Lorentz indices, but the notation is useful for organizing the indices more clearly.

Since the propagator has equally many factors of momenta in the numerator and denominator, it must have originated from a gauge-fixed Lagrangian with a non-local kinetic term $\sim \partial \partial/\Box$. The interaction term is straightforward to obtain using a product of two Levi-Civita tensors. Altogether the double-copy Lagrangian can be written as\begin{equation}\label{eq:cs-double-copy}
\mathcal{L} = \epsilon_{\mu\nu\rho}\epsilon_{\dot{\mu}\dot{\nu}\dot{\rho}}H^{\mu\dot\mu}\frac{\partial^{\nu}\partial^{\dot{\nu}}}{\square}H^{\rho\dot\rho} +
\frac{1}{3}
\epsilon_{\mu\nu\rho}\epsilon_{\dot{\mu}\dot{\nu}\dot{\rho}}H^{\mu\dot\mu}H^{\nu\dot\nu}H^{\rho\dot\rho}\ ,
\end{equation}
where the field $H$ is restricted to Lorenz gauge $\partial_\mu H^{\mu\dot\mu}=0=\partial_{\dot \mu} H^{\mu\dot\mu}$. It is possible that after appropriately removing the gauge fixing, together with introducing auxiliary fields and using field redefinitions, this non-local Lagrangian can be massaged into a more standard one. We will not attempt this exercise here.  
The coupling constant is suppressed in the above, but it is clear that it is dimensionful in 3D (dimensional reduction to 0D gives a marginal theory). Finally, we note that the interaction term is a determinant ${\rm Det}(H)$, suggesting that the theory has some interesting properties that deserve further exploration.

\subsection{Chern-Simons $\otimes$ Chern-Simons in superfield notation}
In the superspace formulation of Chern-Simons theory we can construct a double copy that automatically includes all bosonic fields and ghosts.\footnote{See ref.~\cite{Anastasiou:2018rdx,Borsten:2020xbt,Borsten:2020zgj, Borsten:2021hua} for double copies in Yang-Mills theory that includes the BRST ghosts.} Color-kinematics duality relies on Lorenz gauge $b\Psi=0$, and thus the double copy Lagrangian will be obtained in a gauge-fixed form. 

Let us start by upgrading the kinetic term in~\eqn{eq:cs-double-copy} to superspace 
\begin{equation}\label{eq:super-cs-double-copy-kinetic-term}
\mathcal{L}_{\rm kin} = \int d^3\theta d^3\tilde{\theta}
\ 
\textbf{H} \ 
\frac{\textbf{Q}}{\square}
\ 
\textbf{H} \ ,
\end{equation}
with the bosonic superfield $\textbf{H}= \theta_\mu H^{\mu\dot{\mu}}\tilde{\theta}_{\dot{\mu}}$, and new bosonic differential operator $\textbf{Q}=Q\tilde{Q}$, where $\tilde{Q}=\tilde \theta \cdot \partial$. Thus we have doubled the number of Grassmann-odd parameters by introducing $\tilde{\theta}^{\dot{\mu}}$. We also introduce a second copy of the operator $\tilde{b}= \tilde \bop \cdot \partial$, with $\tilde \bop^{\dot \mu} =\partial/\partial \tilde \theta_{\dot \mu}$. 

Defining the product operator $\textbf{b}\equiv -b\tilde{b}$, one can show that it satisfies 
\begin{equation}
\textbf{b}^2 = 0 \ \ \ \textrm{and}\ \ \ 
\frac{\textbf{b}}{\square}\frac{\textbf{Q}}{\square} f = f -\frac{1}{\square}\tilde{Q}\tilde{b}f - \frac{1}{\square}Qb f - \frac{1}{\square^2}Q\tilde{Q} b\tilde{b}f \ ,
\end{equation}
where $f$ is some test function. Assuming that the test function satisfies the Lorenz gauge conditions, $bf=0=\tilde{b}f$, then the object $\textbf{b}/\square$ is the formal inverse of $\textbf{Q}/\square$ when acting on $f$.  Using this property we may extend the $\textbf{H}$-field Lagrangian to include a complete superfield
\begin{equation}\label{eq:super-cs-double-copy-action}
S = \int d^3x d^3\theta d^3\tilde{\theta}
\ 
\frac{1}{2}\boldsymbol{\Psi} \ 
\frac{\textbf{Q}}{\square}
\ 
\boldsymbol{\Psi} + \frac{1}{3!}\boldsymbol{\Psi}\boldsymbol{\Psi}\boldsymbol{\Psi}\ ,
\end{equation}
where the Grassmann-even superfield $\boldsymbol{\Psi}$ is a double-copy of two Chern-Simons superfields,
\begin{equation}
\boldsymbol{\Psi} = \Psi \otimes \tilde \Psi 
=
( c + \theta_\mu A^\mu + \theta_\mu \theta_\nu {C}^{\mu\nu} + \theta_1\theta_2\theta_3 a) \otimes ( \tilde c + \tilde \theta_{\dot\mu} \tilde A^{\dot \mu} + \tilde \theta_{\dot \mu} \tilde\theta_{\dot \nu} \tilde C^{\dot\mu\dot\nu} + \tilde\theta_{\dot 1}\tilde\theta_{\dot 2}\tilde\theta_{\dot 3} \tilde a)\,, 
\end{equation}
where again we set $a=\tilde a=0$ since they are non-dynamical fields. 

While this action can be used to compute the correlation functions that are obtained from double copying the corresponding Chern-Simons numerators, there is little advantage of using the action for this purpose. Instead, the action is a first step towards elucidating the properties of the resulting double-copy theory. However, to go further one needs to undo the gauge fixing that is implicit in the construction, and find a more elegant action, likely with the non-localities removed.  We leave this for future work.

\subsection{Chern-Simons $\otimes\ (DF)^2$}
An interesting higher-derivative gauge theory that has been used in the double-copy program to construct conformal gravity amplitudes is the $(DF)^2$ theory~\cite{Johansson:2017srf,Johansson:2018ues}. 
In the minimal version of the $(DF)^2$ Lagrangian, there is only the kinetic term 
\begin{equation}
\mathcal{L} =  \frac{1}{g^2} {\rm Tr}\, (D_\mu F^{\mu\nu})^2  \ ,
\end{equation}
where the field strength was defined below \eqn{CSaction}, and $D_\mu$ is the covariant derivative satisfying $[D_\mu,D_\nu]=F_{\mu \nu}$. The non-minimal version of the theory further contains a scalar field $\varphi$ and higher-derivative interaction terms: $F^3$, $\varphi F^2$ and $\varphi^3$, and this theory is intimately connected to the bosonic and heteroic string amplitudes~\cite{Azevedo:2018dgo}.  We will not need the non-minimal interactions here as we will focus on the linearized theory.  

Amplitudes in both the minimal and non-minimal $(DF)^2$ theories have been shown to obey the color-kinematics duality~\cite{Johansson:2017srf,Johansson:2018ues}.  This suggest that we can consider double copying the theories also off shell with Chern-Simons theory. While off-shell BCJ numerators in $(DF)^2$ are not yet known, we only need to use standard cubic Feynman-diagram numerators, since it is sufficient that one copy of numerators are manifestly in BCJ form~\cite{Bern:2010ue}.  The $(DF)^2$ Lagrangian is marginal in 6D and this has the interesting consequence that the double-copy operation ``$(DF)^2 \otimes$'' applied to a theory that is marginal in $D$ dimensions will give back a new theory marginal in the same dimension. For example, $(DF)^2 \otimes {\rm YM}_{\rm 4D}$ gives 4D conformal gravity~\cite{Johansson:2017srf,Johansson:2018ues}. Likewise, we should expect that $(DF)^2 \otimes {\rm CS}_{\rm 3D}$ should give a 3D marginal theory. 

Hence we expect that the resulting double-copy theory should be some diffeomorphism invariant theory, classically conformal in 3D. A likely candidate is Chern-Simons gravity~\cite{Deser1982},
\begin{equation}
S_{\Gamma} = \frac{1}{2\pi}\int d^3x\, \epsilon^{\mu\nu\rho}\left(\Gamma^{\alpha}_{\mu\beta}\partial_\nu \Gamma^{\beta}_{\rho\alpha}  + \frac{2}{3}\Gamma^{\alpha}_{\mu\beta}\Gamma^{\beta}_{\nu\gamma}\Gamma^\gamma_{\rho\alpha} \right) \ ,
\end{equation}
where $\Gamma^\alpha_{\mu\beta}$ are the Christoffel symbols defined in the usual way through the metric field $g_{\mu \nu}=\eta_{\mu \nu}+ h_{\mu \nu}$, and we have suppressed the coupling. The equation of motion is simply ${\cal C}^{\mu\nu}=0$, where the Cotton tensor is given by
\begin{equation}
{\cal C}^{\mu\nu} = \epsilon^{\mu\rho\delta}\nabla_\rho R_\delta^\nu + (\mu\leftrightarrow \nu) \ ,
\end{equation}
where $R_{\mu \nu}$ and $\nabla_\rho$ are the Riemann tensor and covariant derivative, respectively.  
In order to compare with the double-copy theory we again look at the propagator, obtained from the $(DF)^2$ propagator by acting on it with the numerator of the Chern-Simons propagator, 
\begin{equation}
G^{\mu\dot \mu;\nu \dot \nu} \equiv -i \epsilon^{\mu \rho \nu}\partial_\rho  G^{\dot \mu\dot \nu}_{(DF)^2} = \int \frac{d^3p}{(2\pi)^3}
\frac{\epsilon^{\mu \rho \nu}p_\rho }{p^4}
\left(
\eta^{\dot \mu \dot \nu}-\frac{1-\xi}{p^2}p^{\dot \mu}p^{\dot \nu}
\right)e^{ip\cdot x} \ .
\end{equation}
Here, the gauge-fixing parameter $\xi$ of the $(DF)^2$ theory is kept for clarity.
This propagator is the inverse of the kinetic term
\begin{equation} \label{kintermDFCS}
{\cal L}_{\rm kin}=\epsilon^{\mu \nu \rho} H_{\mu \dot{\nu}}\left(
\eta^{\dot \mu \dot \rho} \square  \partial_\nu
+\frac{1-\xi}{\xi}\partial^{\dot{\mu}}\partial^{\dot{\rho}}\partial_\nu
\right) H_{\rho \dot \rho} \ ,
\end{equation}
where $ H_{\mu \dot{\nu}}=  A_{\mu}^{\rm CS} \otimes A^{(DF)^2}_{\dot{\nu}}$. 

Let us decompose the $H_{\mu \dot{\nu}}$ field into irreducible Lorentz representations: $H_{\mu \nu} = h_{((\mu \nu))} + B_{[\mu \nu]} + \eta_{\mu \nu} \phi$, where we temporarily dropped the dot on the indices.  We should be able to associate the symmetric traceless field $h_{\mu \nu}$ with the linear perturbation of the metric. Doing so, we can see that the equations of motion coming from the kinetic term (\ref{kintermDFCS}) indeed imply the vanishing of the Cotton tensor, which to linearized order is
\begin{equation}
{\cal C}^{\mu\nu}\Big|_{\rm linear} = \epsilon^{\mu \alpha \beta}(\partial_\alpha \partial_\gamma \partial^\nu h^\gamma_\beta - \partial_\alpha \square h^\nu_\beta )+ (\mu \leftrightarrow \nu) \ .
\end{equation}
This already gives strong circumstantial evidence that the double copy of Chern-Simons theory with $(DF)^2$ theory gives some version of Chern-Simons gravity, with additional tensor and scalar fields. However, further non-trivial checks at the non-linear order in the fields are desirable, and we leave this task to upcoming work.

\subsection{Chern-Simons $\otimes$ Yang-Mills} 
In order to double-copy Chern-Simons with Yang-Mills theory, we need a cubic action for the theory. We will not use a formulation that manifestly satisfies the color-kinematics duality, since it is sufficient that the Chern-Simons side of the double copy satisfies the duality. 
We will use the following cubic Yang-Mills Lagrangian:
\begin{equation}
\mathcal{L}_{\rm YM} = -\frac{1}{g^2} \textrm{Tr}\Big(2\partial^\mu A^\nu \partial_{[\mu} A_{\nu]}
+ 2 [A^{\mu},A^{\nu}]\partial_\mu A_\nu
+ [A_{\mu},A_{\nu}]\partial_\rho B^{\rho \mu\nu}
+ \frac{1}{2}B_{\rho \mu\nu} \Box B^{\rho \mu\nu} \Big) \ .
\end{equation}
If needed, the $B$ field can be integrated out $B_{\rho \mu\nu} = \frac{\partial_\rho}{\Box}[A_\mu, A_\nu]$, to recover the standard Yang-Mills Lagrangian, ${\cal L}=-1/2g^2\,\textrm{Tr}\, (F_{\mu \nu})^2$.

We now have two reducible fields $H_{\mu\dot\mu}\equiv A^{\rm CS}_{\mu}\otimes A^{\rm YM}_{\dot\mu}$, and $K_{\mu \dot\rho\dot \mu\dot\nu}\equiv A^{\rm CS}_{\mu}\otimes B^{\rm YM}_{\dot\rho \dot\mu\dot\nu}$ from the double copy with Chern-Simons theory. When both the left and right side gauge fields are in Lorenz gauge, the propagator for the $H$ field takes the form
\begin{equation}
G_H^{\mu\dot{\mu},\nu\dot{\nu}}(x) = \int \frac{d^3p}{(2\pi)^3} \frac{
\epsilon^{\mu\alpha\nu}p_\alpha 
\eta^{\dot{\mu}\dot{\nu}}
}{p^2}e^{ip\cdot x} \ .
\end{equation}
Similarly, the propagator for the $K$ field has to take the form
\begin{equation}
G_K^{\mu\dot{\alpha}\dot\beta\dot\gamma, \nu\dot \rho\dot\mu\dot\nu}(x) = -2\int \frac{d^3p}{(2\pi)^3} \frac{
\epsilon^{\mu\alpha\nu}p_\alpha 
\eta^{\dot\alpha\dot{\rho}}\eta^{\dot\beta\dot\mu}\eta^{\dot\gamma\dot\nu}
}{p^2}e^{ip\cdot x} \ .
\end{equation} 
Because the propagators are simply the same as in Chern-Simons theory dressed with additional trivial $\eta^{\dot \mu \dot\nu}$ factors, the kinetic terms for the $H$ and $K$ fields will be very similar to the Chern-Simons kinetic term. The interaction terms are straightforward to obtain from the double copy, and the resulting Lagrangian is
\begin{eqnarray}
\mathcal{L} &=& \frac{1}{2}H^{\mu\dot{\mu}}\epsilon_{\mu\nu\rho}\partial^\nu H^{\rho}_{\ \dot{\mu}}
+
\frac{1} {6}
H^{\mu\dot{\mu}}H^{\nu\dot{\nu}}\epsilon_{\mu\nu\rho}\partial_{\dot{\mu}} H^{\rho}_{\ \dot{\nu}} 
\nonumber \\ 
&&\null
-\frac{1}{4}K^{\mu\dot \sigma \dot{\alpha}\dot{\beta}}\epsilon_{\mu\nu\rho}\partial^\nu K^{\rho}_{\ \dot \sigma \dot{\alpha}\dot{\beta}} + \frac{1}{2}H^{\mu\dot{\mu}}H^{\nu\dot{\nu}}\partial^{\dot \sigma} K^\rho_{\  {\dot \sigma}\dot{\mu}\dot{\nu}}\epsilon_{\mu\nu\rho} \ .
\end{eqnarray}
As usual in the double copy Lagrangians, we have suppressed the coupling constant, which is dimensionful in 3D, but becomes marginal if the theory is dimensionally reduced to 1D. The presented Lagrangian is manifestly local;  the $H$ field has inherited the gauge condition $\partial_{\dot \mu}H^{\mu\dot{\mu}}=0$ from the Yang-Mills vector, but the $K$-field can be assumed to be unconstrained. It is likely that this Lagrangian can be further massaged into a more elegant or recognizable form, which we will not attempt here.

\section{Chern-Simons-matter theories}
\label{Section6}
We will now work out new examples of Chern-Simons-matter theories that have on-shell amplitudes that obey color-kinematics duality.\footnote{Note that we here do not consider the exotic 3-Lie algebra version of color-kinematics duality, which was studied for BLG and ABJM theories in refs.~\cite{Bargheer:2012gv,Huang:2012wr,Huang:2013kca}. The new examples of Chern-Simons-matter theories presented here, obey the standard version of color-kinematics duality.}
Amplitudes can be defined for Chern-Simons-matter theories, involving an even number of external states, typically in some matter representation of the gauge group. Here we explore on-shell color-kinematics duality for matter only in the adjoint representation of $SU(N_c)$. We remind the reader that color-ordered amplitudes are given with respect to rescaled structure constants $\tilde{f}^{abc} = i\sqrt{2}f^{abc} $. The relevant Feynman rules can be found in appendix \ref{Feynmanrules}.

\subsection{Scalar matter}

Consider adding adjoint complex scalars to the Chern-Simons Lagrangian, with kinetic term 
\begin{equation}
\mathcal{L}_{\rm kin} =
2\,{\rm Tr}\big( |D_\mu\phi^{i}|^2\big)  \ ,
\end{equation}
and with flavor indices $i=1, \ldots, N_f$ belonging to $U(N_f)$. With this Lagrangian term, we find that the relevant four-scalar Chern-Simons tree amplitudes are
\begin{eqnarray}
A_4(\phi_1\bar{\phi}_2\phi_3\bar{\phi}_4)& =& 2\epsilon^{p_1 p_2 p_3}\left( 
(-1)^{|\phi|}\frac{\delta^{i_1 \bar \imath_4}\delta^{i_3 \bar \imath_2}}{s_{23}}
-\frac{\delta^{i_1 \bar \imath_2}\delta^{i_3 \bar \imath_4}}{s_{12}}
\right) \ ,
\nonumber \\
A_4(\phi_1\bar{\phi}_2\bar{\phi}_4\phi_3)& =& 2\epsilon^{p_1 p_2 p_3}\frac{\delta^{i_1 \bar \imath_2}\delta^{i_3 \bar \imath_4}}{s_{12}} \ .
\end{eqnarray}
Here we are anticipating that the scalars may carry a nontrivial statistical phase under exchange of particles, and this is captured by the factor $(-1)^{|\phi|}$. The four-point BCJ relation that we want to impose is $s_{24}A_4(1,2,4,3) = s_{14}A_4(1,2,3,4)$, and this gives the equation
\begin{eqnarray}
(-1)^{|\phi|}\delta^{i_1 \bar \imath_4}\delta^{i_3 \bar \imath_2} + \delta^{i_1 \bar \imath_2}\delta^{i_3 \bar \imath_4} = 0 \ ,
\end{eqnarray}
which has only one solution $|\phi|=1$ and $\delta^{i \bar \jmath}=1$. This means the scalars are anti-commuting fields, and the flavor group is $U(1)$, hence we drop the flavor indices.  After plugging in the solution, the four-point amplitudes in our Chern-Simons-matter theory are
\begin{equation}
A_4(\phi_1\bar{\phi}_2\phi_3\bar{\phi}_4) = -2\epsilon^{p_1 p_2 p_3}\left( 
\frac{1}{s_{23}}
+\frac{1}{s_{12}}
\right) \,, ~~~~A_4(\phi_1\bar{\phi}_2\bar{\phi}_4\phi_3) = \frac{2\epsilon^{p_1 p_2 p_3}}{s_{12}} \ .
\end{equation}

Now let us consider higher-multiplicity amplitudes. We add to the theory a six-scalar contact interaction respecting the $U(1)$ symmetry, giving the marginal matter Lagrangian 
\begin{equation} \label{scalarL}
\mathcal{L}_{\phi} =
(D_\mu \bar{\phi})^a (D^\mu \phi)^a - \lambda g^4 \phi^{a}\bar{\phi}^{b}\bar{\phi}^{c}\phi^{d}\phi^{e}\bar{\phi}^{h}{f}^{a b x}{f}^{x c y}{f}^{y d z}{f}^{z e h}\ .
\end{equation}
All other six-point interactions can be related to this one via identities of the structure constants.  Here $g$ is the gauge coupling constant, which for convenience we set to $g=1$ in the below calculations, and $\lambda$ is a fudge factor that will be fixed by demanding that the theory obeys the BCJ relations at six points and above.

The simplest tree amplitude at six points is the color-ordering $A_6(\phi_1\phi_2\phi_3 \bar{\phi}_4\bar{\phi}_5\bar{\phi}_6) $, which we depict as a disk diagram with matter lines stretching between boundary points. All contributing Feynman diagrams can be obtained by dressing the disk diagram with Chern-Simons vector lines, in all possible ways that respect planarity. The result is
\begin{eqnarray}
A_6(\phi_1\phi_2\phi_3 \bar{\phi}_4\bar{\phi}_5\bar{\phi}_6) &=&
\begin{tikzpicture}[baseline={(0, -0.1cm)}]
\draw[color=gray] (1,0) arc [start angle=0,end angle=360,radius=1cm] ;
\draw[thick] (-0.5,0.866) .. controls (-0.2,0.5) and (0.2,0.5) .. (0.5,0.866);
\draw[thick] (-1,0) -- (1,0);
\draw[thick] (-0.5,-0.866) .. controls (-0.2,-0.5) and (0.2,-0.5) .. (0.5,-0.866);
\node at (-1.2,0) {$2$};\node at (1.2,0) {$\bar{5}$};
\node at (-0.65,1.1) {$3$};\node at  (0.65,1.1) {$\bar{4}$};
\node at (-0.65,-1.1) {$1$};\node at  (0.65,-1.1) {$\bar{6}$};
\end{tikzpicture} \nn
\\
&=&
 4i\frac{ \epsilon^{p_2 p_3 p_4}\epsilon^{p_1 p_5 p_6}}{s_{16}s_{34} s_{234}} +4i \frac{ \epsilon^{p_3 p_4 p_5}\epsilon^{p_1 p_2 p_6}}{s_{16}s_{34}s_{345}} 2i \frac{ \epsilon_{p_3 p_4 \mu}\epsilon^{\mu p_6 p_1}}{s_{34}s_{16}} - \frac{i}{2} \lambda \ ,
\end{eqnarray}
where $s_{ij\ldots k} = (p_i+p_j+\ldots+p_k)^2$.

The next simplest amplitude is $A_6(\phi_1\phi_2\bar{\phi}_3\phi_4\bar{\phi}_5\bar{\phi}_6)$, which can be written in terms of the amplitude above plus an additional disk diagram,
\begin{eqnarray}
A_6(\phi_1\phi_2\bar{\phi}_3\phi_4\bar{\phi}_5\bar{\phi}_6) &=&
\begin{tikzpicture}[baseline={(0, -0.1cm)}]
\draw[color=gray] (1,0) arc [start angle=0,end angle=360,radius=1cm] ;
\draw[thick] (-0.5,0.866) .. controls (-0.2,0.5) and (0.2,0.5) .. (0.5,0.866);
\draw[thick] (-1,0) -- (1,0);
\draw[thick] (-0.5,-0.866) .. controls (-0.2,-0.5) and (0.2,-0.5) .. (0.5,-0.866);

\node at (-1.2,0) {$2$};\node at (1.2,0) {$\bar{5}$};
\node at (-0.65,1.1) {$\bar{3}$};\node at  (0.65,1.1) {$4$};
\node at (-0.65,-1.1) {$1$};\node at  (0.65,-1.1) {$\bar{6}$};
\end{tikzpicture}
+
\begin{tikzpicture}[baseline={(0, -0.1cm)}]
\draw[color=gray] (1,0) arc [start angle=0,end angle=360,radius=1cm] ;
\draw[thick]  (-1,0) .. controls (-0.55,0.1) and (-0.4,0.4) .. (-0.5,0.866);
\draw[thick]  (1,0) .. controls (0.55,0.1) and (0.4,0.4) .. (0.5,0.866);
\draw[thick] (-0.5,-0.866) .. controls (-0.2,-0.5) and (0.2,-0.5) .. (0.5,-0.866);

\node at (-1.2,0) {$2$};\node at (1.2,0) {$\bar{5}$};
\node at (-0.65,1.1) {$\bar{3}$};\node at  (0.65,1.1) {$4$};
\node at (-0.65,-1.1) {$1$};\node at  (0.65,-1.1) {$\bar{6}$};
\end{tikzpicture}
\\
&=&
A(\phi_1\phi_2\phi_3\bar{\phi}_4\bar{\phi}_5\bar{\phi}_6) + \frac{2i \epsilon^{\mu\nu\rho}\epsilon_{\mu p_2 p_3}\epsilon_{\nu p_4 p_5}\epsilon_{\rho p_1 p_6}}{s_{16}s_{23}s_{45}} + \frac{3i}{4}\lambda \\
&&\null
+ i\left( \frac{\epsilon_{p_6 p_1 \mu}\epsilon^{\mu p_5 p_4}}{s_{16}s_{45}} + \frac{4 \epsilon_{p_1 p_2 p_6}\epsilon_{p_3 p_4 p_5}}{s_{16}s_{45}s_{345}} +\textrm{cyclic}(\{12\},\{34\},\{56\}) \right)
\, .\nonumber 
\end{eqnarray}
Here the last parenthesis should include all three permutations of the cyclic orbit $\{1,2\}\to\{3,4\}\to \{5,6\}$.

From these two partial amplitudes all other color orderings are obtained through permutations, and thus it is possible to test all six-point BCJ relations~(\ref{BCJrel}). After doing so, one fins a single constraint on the fudge parameter  $\lambda = 1$. Further checks of the BCJ relations up to eight points give no additional constraints, confirming that the matter Lagrangian~(\ref{scalarL}) combined with the pure Chern-Simons action~(\ref{CSaction}) gives a theory that obeys on-shell color-kinematics duality.

\subsection{Fermion matter}
Chern-Simons theory can be coupled to a two-component Dirac fermion, which we take to be in the adjoint representation of the gauge group. The matter Lagrangian is the standard kinetic term,
\begin{equation} \label{fermionKin}
\mathcal{L}_{\rm kin}=2 \, {\rm Tr }\, \bar{\psi}\,i\slashed{D}\psi \ ,
\end{equation}
and no other terms with dimensionless couplings are permitted. Since the fermion is complex it is charged under a $U(1)$ flavor symmetry, and it describes two degrees of freedom. 

It is convenient to make use of the 3D spinor helicity notation, for which our conventions are collected in Appendix \ref{spinorconventions}. As there is no chirality in 3D, we have only one type of on-shell spinor that we take to satisfy
\begin{equation}
|p \rangle \langle p| = - p^\mu \gamma_\mu\,, ~~~~ p^2=0\ .
\end{equation}
In the theory with only fermion matter, there exists no marginal higher-point interactions that can be added to the Lagrangian, and thus the BCJ relations can at best be checked to hold, there is no tuning of parameters permitted. 

The four-fermion tree amplitudes can be computed as
\begin{equation}
A_4(\psi_1\bar{\psi}_2\psi_3\bar{\psi}_4) =
\frac{\langle 1 | \gamma^\mu | 2\rangle \langle 3 | \gamma^\nu | 4\rangle \epsilon_{\mu p_{12}\nu} }{2s_{12}}
+\frac{\langle 1 | \gamma^\mu | 4\rangle \langle 2 | \gamma^\nu |3\rangle \epsilon_{\mu p_{14}\nu} }{2s_{14}} =
\frac{\langle 1 3 \rangle ^3}{\langle 2 1 \rangle \langle 14\rangle } \ , 
\end{equation}
and the other distinct ordering, with fewer intermediate steps, is given by
\begin{equation}
A_4(\psi_1\bar{\psi}_2\bar{\psi}_4\psi_3) = \frac{\langle 4 2 \rangle \langle 2 3 \rangle}{\langle 2 1 \rangle } \ .
\end{equation}
These automatically obey the  BCJ relation $s_{24} A_4(1,2,4,3) = s_{14}A_4(1,2,3,4)$, which confirms that color-kinematics duality is obeyed up to this multiplicity.
  
We note that under exchange of identical particles $2\leftrightarrow 4$ (or $1\leftrightarrow 3$) the four-point amplitude is even, which implies that the spinors behave as commuting fields. Thus for the particular setup we have landed on, with minimal amounts of complex matter (scalar or fermion) in the adjoint representation, we see that the BCJ-satisfying Chern-Simons theories give opposite-statistics matter. While, in principle, the spin-statistics theorem does not hold in 3D, to our knowledge one typically chooses to work with standard spin-statistics when describing the fundamental matter degrees of freedom. We will not need to dive deeper into this topic, as the unusual statistics of our gauge-theory matter becomes the standard statistics after the double copy is applied. 

With the same matter Lagrangian it is straightforward to compute the six-fermion partial amplitudes, for instance the simplest one is
\begin{equation}
A_6(\psi_1\psi_2\psi_3\bar{\psi}_4\bar{\psi}_5\bar{\psi}_6) = -\frac{i}{4}\frac{\langle 34\rangle \langle 16\rangle \langle 2|(p_3-p_4)p_{234}(p_1 -p_6)|5\rangle}{s_{34}s_{16}s_{234}} + (3,4)\leftrightarrow (1,6) \ .
\end{equation}
We checked that this amplitude together with the other six-fermion partial amplitudes obeys all the BCJ relations at six points. Furthermore, we computed all the eight-fermion partial amplitudes from the Feynman rules, finding that they obey the corresponding BCJ relations. This confirms that simply adding the fermion kinetic term~(\ref{fermionKin}) to the Chern-Simons action~(\ref{CSaction}) gives a theory that obeys on-shell color-kinematics duality.

\subsection{Scalars and fermions: $\mathcal{N}=4$ Chern-Simons matter}\label{Neq2}
It is natural to attempt to combine the scalar and fermion theories. Aside from the matter Lagrangians already presented, one finds that there are two new possible marginal couplings between the scalars and fermions of the form $\bar{\psi}\psi \bar\phi \phi$. Therefore, to fix their coefficients it is sufficient to study four-point amplitudes. The two additional independent interaction terms are
\begin{eqnarray}
\mathcal{L}_{\rm int}  &=&
i g^2\bar{\psi}^a \psi^b \bar{\phi}^c\phi^d (\alpha f^{acx}f^{xbd} 
+
\beta f^{adx}f^{xbc}) \ ,
\end{eqnarray}
where the free coefficients $\alpha$ and $\beta$ are to be determined through the BCJ relations. 
Next, using the Feynman rules in appendix \ref{Feynmanrules}, one can obtain the amplitudes
\be
A(\psi_1 \phi_2 \bar{\phi}_3 \bar{\psi}_4) ~ = ~
\begin{tikzpicture}[baseline={(0, -0.1cm)}]
\draw[thick,snake it] (0,-0.33) -- (0,0.43);
\draw[thick] (0.5,-0.6) -- (0,-0.33);
\draw[thick] (-0.5,-0.6) -- (0,-0.33);
\draw[thick,dashed] (0.5,0.7) -- (0,0.43);
\draw[thick,dashed] (-0.5,0.7) -- (0,0.43);
\node at (-0.55*1.25,-0.6*1.25) {$1$};
\node at (0.55*1.25,-0.6*1.25) {$\bar{4}$};
\node at (0.55*1.25,0.7*1.25) {$\bar{3}$};
\node at (-0.55*1.25,0.7*1.25) {$2$};
\end{tikzpicture} 
+
\begin{tikzpicture}[baseline={(0, -0.1cm)}]
\draw[thick,dashed] (-0.5,0.5) -- (0,0);
\draw[thick] (-0.5,-0.5) -- (0,0);
\draw[thick,dashed] (0.5,0.5) -- (0,0);
\draw[thick] (0.5,-0.5) -- (0,0);
\node at (-0.5*1.25,-0.55*1.25) {$1$};
\node at (-0.5*1.25,0.55*1.25) {$2$};
\node at (0.5*1.25,0.55*1.25) {$\bar{3}$};
\node at (0.5*1.25,-0.55*1.25) {$\bar{4}$};
\end{tikzpicture} 
 ~ = ~
 -\frac{1}{2s_{14}}\langle 1 4\rangle (s_{12}-s_{13}+\alpha s_{14})  \ ,
\ee
and
\be
A(\psi_1 \phi_2 \bar{\psi}_4 \bar{\phi}_3) ~ = ~ 
\begin{tikzpicture}[baseline={(0, -0.1cm)}]
\draw[thick,dashed] (-0.5,0.5) -- (0.5,-0.5);
\draw[thick] (-0.5,-0.5) -- (0.5,0.5);
\node at (-0.5*1.25,-0.55*1.25) {$1$};
\node at (-0.5*1.25,0.55*1.25) {$2$};
\node at (0.5*1.25,0.55*1.25) {$\bar{4}$};
\node at (0.5*1.25,-0.55*1.25) {$\bar{3}$};
\end{tikzpicture}
~ = ~
\frac{1}{2}(\alpha +\beta) \langle 14 \rangle  \ .
\ee
Demanding that the BCJ relation $s_{14}A(1234) = s_{24}A(1243)$ holds, leads to the constraint
\begin{equation}
(s_{12} - s_{13} + \alpha s_{23}) =  (\alpha+\beta) s_{24} \ ,
\end{equation}
with the unique solution $\alpha =\beta = 1$.
No additional constraints are found at four points, nor at higher multiplicity.
Putting all of this together, we find the following Chern-Simons-matter Lagrangian that obeys on-shell color-kinematics duality:
\begin{eqnarray} \label{Neq4Lagr}
\mathcal{L}_{\mathcal{N}=4} &=& \frac{\epsilon_{\mu\nu\rho}}{2}\left(
 A^{a \mu} \partial^\nu A^{a \rho} - \frac{g}{3}f^{abc}A^{a\mu} A^{b\nu} A^{c\rho}\right) +
(D_\mu \bar{\phi})^a(D^\mu \phi)^a +
i\bar{\psi}^a(\slashed{D}\psi )^a
  \\
&&\null
+
i g^2\bar{\psi}^a \psi^b \bar{\phi}^c\phi^d (f^{acx}f^{xbd} 
+
f^{adx}f^{xbc})
-
g^4 \phi^{a}\bar{\phi}^{b}\bar{\phi}^{c}\phi^{d}\phi^{e}\bar{\phi}^{h}{f}^{a b x}{f}^{x c y}{f}^{y d z}{f}^{z e h}\nonumber 
 \ ,
\end{eqnarray}
where the gauge coupling and color indices are explicitly shown. 

We can now demonstrate that the Lagrangian (\ref{Neq4Lagr}) is very special, in that it possesses $\mathcal{N}=4$ supersymmetry. To show this we need to first identify the R-symmetry, which can be as large as $SO(\mathcal{N})$ in 3D. In our construction, we have only made manifest a $U(1)\times U(1)$ symmetry, where the scalar and fermion carry distinct charges. Of course, $U(1)\times U(1)$ has the same rank as $SU(2)\times SU(2) \sim SO(4)$, which motivates the introduction of two pseudo-real doublet fields to make this symmetry manifest,
\be
\phi_\alpha = (\bar{\phi},\phi)\,,
~~~~~~~ \psi_{\dot{\alpha}} = (\bar{\psi},\psi)\,,
\ee 
where the $SU(2)$ indices can be raised (and lowered) with a Levi-Civita symbol $\epsilon^{\alpha \beta}$, and dotted indices indicate the right $SU(2)$ factor. Thus the scalars transform as $(\textbf{2},\textbf{1})$ and fermions as $(\textbf{1},\textbf{2})$ in the R-symmetry group $SO(4)$.

Next, we present the supersymmetry transformations that we find to leave the Lagrangian (\ref{Neq4Lagr})  invariant.  The supersymmetry transformation parameter $\xi^{\alpha\dot\alpha}$ is a Grassman-odd Lorentz spinor and $SO(4)$ vector (bi-spinor). Defining the composite field $(\tau^c)_{\alpha\beta}\equiv\phi^a_\alpha\phi^b_\beta {f}^{abc}$, the supersymmetry transformations relevant to the Lagrangian (\ref{Neq4Lagr}) are
\begin{eqnarray}
\delta \phi_\alpha &=& \bar{\xi}_\alpha^{\ \dot\alpha}\psi_{\dot\alpha}\ , \nn \\
\delta A_\mu^a &=&  g\bar{\xi}^{\alpha\dot\alpha}\gamma_\mu\psi^b_{\dot\alpha}\phi^c_\alpha {f}^{abc} \ ,\\
\delta \psi^{a}_{\dot\alpha} 
&=&
i (\slashed{D} \phi^{\alpha})^a\xi_{\alpha \dot\alpha}   -\frac{i g^2}{3}{f}^{abc}\phi_\beta^b
(\tau^c)^{\beta \alpha }
\xi_{\alpha \dot\alpha} \ , \nn
\end{eqnarray}
where the bared spinors denote Dirac conjugates. 

We can now study the amplitudes in a manifestly supersymmetric manner. For this purpose, we define the on-shell half-hypermultiplet
\begin{equation}
\HYPER = \psi + \eta_\alpha \phi^\alpha + \eta_1 \eta_2 \bar{\psi} \ ,
\end{equation}
where the $\eta_\alpha$ variables are Grassmann odd auxiliary parameters that transform as $(\textbf{2},\textbf{1})$ in $SO(4)$, thus making one factor of $SU(2)$ manifest.\footnote{By a half-Fourier transform over the $\eta$ variables the other $SU(2)$ factor can be made manifest in the half-hypermultiplet.} The fields are here best interpreted as the asymptotic on-shell solutions to field equations, but for identification purposes we use the same notation as for the off-shell fields. 

In terms of this on-shell superfield, the four-point amplitudes can be written as
\begin{equation} \label{superAmp}
A_4(\HYPER _1\HYPER _2\HYPER _3\HYPER _4) = \delta^4(Q) \frac{\langle 13\rangle \langle 24\rangle }{\langle 1 2\rangle \langle 2 3\rangle \langle 3 1\rangle} \ ,
\end{equation}
where the supercharge (or supermomentum) is defined as
\begin{equation}
Q^\alpha = \sum_{i=1}^{n} |i\rangle \eta_i^\alpha\,,
\end{equation}
and the Grassmann-valued delta function then becomes
\begin{equation}
\delta^4(Q) = \prod_{\alpha=1}^2 \sum_{i,j=1}^{n} \langle i j\rangle \eta_i^\alpha \eta_j^\alpha\,,
\end{equation}
where $n$ is the number of particles; at four points $n=4$. It is easy to confirm that the super-amplitude~(\ref{superAmp}) obeys the BCJ relations, and that it is even under cyclic permutations. 

We now make the observation that this purely-adjoint $\mathcal{N}=4$ Chern-Simons-matter theory can be related to the more standard Chern-Simons-matter theories with bi-fundamental matter in the product gauge group $SU(\tilde N_c) \times SU(\tilde N_c)$. We will not attempt to match the gauge group details, but we note that the color-ordered tree-level amplitudes that the Lagrangian (\ref{Neq4Lagr}) computes are in complete agreement with the corresponding color-ordered amplitudes of the Gaiotto-Witten $\mathcal{N}=4$ supersymmetric Chern-Simons-matter theory~\citep{Gaiotto:2008sd,Hosomichi:2008jd}. We have explicitly checked this through multiplicity eight, by matching to $\mathcal{N}=6$ ABJM amplitudes with supersymmetry truncated to $\mathcal{N}=4$. It is interesting to note that the partial amplitudes of $\mathcal{N}=6$ ABJM theory does not obey the BCJ relations~(\ref{BCJrel}), but upon truncating the $\mathcal{N}=6$ superfield to the one of $\mathcal{N}=4$ Gaiotto-Witten theory, the relations emerge. Thus the $\mathcal{N}=4$ Chern-Simons-matter theory that we studied here appears to exhibit the maximal supersymmetry that is permitted by the BCJ relations~(\ref{BCJrel}). This observation will also be confirmed in the next section when we study the double copy theory, which also exhibits maximal supersymmetry.

\subsection{Double copy: ${\cal N}=8$ Dirac-Born-Infeld theory}
Having constructed several matter theories that obey the BCJ relations, we can now proceed to study the double copies. We begin with the double-copy of the Chern-Simons-scalar theory~(\ref{scalarL}). We find that the double copy can be matched with the dimensional reduction of 6D Born-Infeld theory, which from the 3D perspective is a DBI theory. 

We start from the 6D abelian field strength and metric (in mostly minus signature), and write them in terms of 3D quantities,
\begin{equation}
F_{MN} = \left(\begin{matrix}
F_{\mu \nu} & -\partial_\mu \phi^i \\
\partial_\nu \phi^j & 0 
\end{matrix}\right)\,,~~
\eta_{MN} = \left(\begin{matrix}
\eta_{\mu \nu} & 0\\
0 & -\delta^{ij}
\end{matrix}\right)\ .
\end{equation}
Then the 6D Born-Infeld Lagrangian becomes through dimensional reduction the following DBI theory 
\begin{equation}
{\cal L}_{\rm DBI} = \sqrt{-{\rm det}(\eta_{MN} +\alpha F_{MN})} = \sqrt{{\rm det}(\eta_{\mu \nu} +\alpha F_{\mu \nu} -\alpha^2 \partial_\mu \phi^i  \partial_\nu \phi^i  )}  
 \ ,
\end{equation}
which follows from Schur's determinant identity. In order to simplify the discussion, we will set the coupling constant to unity $\alpha=1$ hereafter. Furthermore, one can dualize the 3D field strength $F_{\mu \nu}=i \epsilon_{\mu \nu \rho}\partial^\rho \phi^0$, and obtain a Lagrangian in terms of a singlet scalar $\phi^0$ and a $SO(3)$ triplet scalar $ \phi^i $ \cite{Schmidhuber:1996fy,Townsend:1995af,polchinski_1998},
\begin{eqnarray}
{\cal L} _{\rm DBI} &=& \sqrt{{\rm det}(\eta_{\mu \nu} +i \epsilon_{\mu \nu \rho}\partial^\rho \phi^0  -\partial_\mu \phi^i  \partial_\nu \phi^i  )} \\
&=& \Bigg(
1-(\partial_\mu \phi^0 )^2-(\partial_\mu \phi^i )^2 
+\frac{1}{2}\partial_\nu \phi^i\partial^\nu\phi^i \partial_\rho \phi^j \partial^\rho \phi^j
-\frac{1}{2}\partial_\nu \phi^i\partial_\rho\phi^i \partial^\nu \phi^j \partial^\rho \phi^j \nonumber \\
&&\null
+\partial_\mu \phi^0 \partial_\nu\phi^0 \partial^\mu\phi^i\partial^\nu\phi^i
-\frac{1}{6}(\partial_\mu \phi^i\partial^\mu\phi^i)^3
+\frac{1}{2} \partial_\mu \phi^i \partial^\mu\phi^i \partial_\nu\phi^j \partial_\rho\phi^j \partial^\nu\phi^k\partial^\rho\phi^k \nonumber\\
&&\null
-\frac{1}{3}\partial_\mu\phi^i\partial^\rho\phi^i \partial_\nu\phi^j \partial^\mu\phi^j \partial_\rho\phi^k\partial^\nu\phi^k
\Bigg)^{-1/2} \ .
\end{eqnarray}
Inside the square root the mixing terms are up to 6th order in the scalar fields, and have  the property that they always involve a product of scalars of different flavor. Hence the mixing terms vanish if the theory is truncated to a single scalar. While the Lagrangian only has manifest $SO(3)$ symmetry, one finds an effective $SO(4)$ symmetry on shell~\cite{Schmidhuber:1996fy,Townsend:1995af,polchinski_1998}, and we explicitly confirmed this for four- and six-point amplitudes.

The $SO(4)$ symmetry can be mapped via the double copy to a $SU(2)\times SU(2)$ symmetry, where the Chern-Simons scalars on each side of the double copy transform as doublets. Indeed, we can make the identification of states via the double copy,
\begin{equation}
\phi\otimes\phi = \phi^0 + i\phi^1 \ , \ \bar{\phi}\otimes\bar{\phi} = \phi^0 -i\phi^1 \ ,\ \phi\otimes\bar{\phi} = \phi^2 + i\phi^3 \ , \ \ \bar{\phi}\otimes\phi = \phi^2 - i\phi^3 \ ,
\end{equation}
where the scalars on the left hand side of the equations denote the two copies of Chern-Simons scalars, and the right-hand-side scalars are those of the DBI theory. This map is simply the manifestation of the familiar $SO(4)$ group-theory identity $(\textbf{1},\textbf{2})\otimes (\textbf{2},\textbf{1})=\textbf{4}$ that can be implemented using Pauli matrices. 

Recall that there is an alternative double-copy construction~\cite{Cachazo:2014xea} of the same DBI theory discussed here, it has the form
\be
{\rm YM}_{\rm 6D} \otimes {\rm NLSM} =({\rm YM}_{\rm 3D} + \phi^i)\otimes {\rm NLSM} = {\rm DBI}\,. 
\ee
The relevant non-linear-sigma-model (NLSM) has a Lagrangian ${\cal L}= {\rm Tr} \, \partial_\mu U^\dagger \partial^\mu U$, where $U$ is a unitary matrix, and it obeys color-kinematics duality~\cite{Chen:2013fya}.  This alternative approach was convenient for cross-checking our results. 

Since the Chern-Simons-scalar theory we studied above is simply a consistent truncation of the  supersymmetric ${\cal N}=4$ Chern-Simons theory, we should expect that the double copy of the latter theory with itself to be a supersymmetric extension of the DBI theory. Indeed, the unique DBI theory that agrees with the expected symmetries and spectrum is the maximally supersymmetric theory, hence we propose the new double-copy relation
\be
 ( {\cal N}=4~{\rm CSm}) \otimes ( {\cal N}=4~{\rm CSm}) ~=~  ( {\cal N}=8~{\rm DBI})\,,
\ee
where ${\cal N}=4$~CSm theory refers to the Lagrangian~(\ref{Neq4Lagr}). 

We now briefly review the maximally supersymmetric 3D ${\cal N}=8$ DBI theory. It is convenient to formulate it in terms of its maximal uplift to 10D, where the action reads~\cite{Aganagic:1996nn}
\begin{equation}
\mathcal{L}_\text{susy-DBI} = \sqrt{
-\textrm{det}(
	\eta_{MN} +  F_{MN} -2 \bar{\chi} \Gamma_M \partial_N \chi + \bar{\chi}\Gamma^P\partial_M \chi \, \bar{\chi} \Gamma_P \partial_N \chi
)
} \ ,
\end{equation}
where $\chi$ are Majorana-Weyl fermions. 
In order to dimensionally reduce this, we give a name to the 10-by-10 matrix that appeared above, 
\begin{equation}
M_{MN} \equiv 
	\eta_{MN} + F_{MN} -2 \bar{\chi} \Gamma_M \partial_N \chi + \bar{\chi}\Gamma^P\partial_M \chi \bar{\chi} \Gamma_P \partial_N \chi \ .
\end{equation}
Its dimensional reduction $SO(1,9)\rightarrow SO(1,2) \times SO(7)$ can be written as
\begin{equation}
M_{MN}=\begin{pmatrix}
M_{\mu\nu} && -\partial_\mu \phi^i  \\
\partial_\nu \phi^j - 2 \bar{\chi} \Gamma^j \partial_\nu \chi && -\delta^{ij} 
\end{pmatrix} \ ,
\end{equation}
where the seven scalars are identified as $\phi^i \equiv A^i$.  
Using Schurs identity as before, we have that the Lagrangian can be written as
\begin{equation}
{\cal L}_{{\cal N}=8\,{\rm DBI}} ~=~ 
\sqrt{{\rm det}(
M_{\mu\nu} - \partial_\mu \phi^i \partial_\nu\phi^i + 2 \partial_\mu \phi^i \bar{\chi}\Gamma^i\partial_\nu \chi
)} \ .
\end{equation}
The Lagrangian has manifest $SO(7)$ R-symmetry, but one can confirm that amplitudes computed from this Lagrangian will exhibit a $SO(8)$ symmetry~\cite{Townsend:1995af,Elvang:2020kuj}, where both the fermions and scalars transform as $\textbf{8}$s, consistent with triality. We can confirm that the double copy $( {\cal N}=4~{\rm CSm})^2$ will exhibit a symmetry larger than $SO(7)$ since the double copy will manifest a group of larger rank, namely $SO(4)\times SO(4) \subset SO(8)$. In fact, explicit calculations confirm that the double copy realizes the full $SO(8)$ R-symmetry. 

Consider the four-point amplitude, using the KLT formula the double copy becomes
\begin{equation}
M_4^{{\cal N}=8}(\mathcal{V}_1\mathcal{V}_2\mathcal{V}_3\mathcal{V}_4) = s_{12}A_4^{{\cal N}=4}(\HYPER_1\HYPER_2\HYPER_3\HYPER_4) A_4^{{\cal N}=4}(\widetilde\HYPER_1 \widetilde \HYPER_2\widetilde\HYPER_4\widetilde\HYPER_3) = \delta^8(Q)\,,
\end{equation}
where the ${\cal N}=8$ ``vector'' superfield is defined as
\begin{equation}
\mathcal{V}^{\mathcal{N}=8} = \HYPER(\eta)\otimes\widetilde\HYPER(\tilde \eta) = A_+ + \eta_A\psi^A_+ + \frac{1}{2} \eta_A \eta_B\phi^{AB} + \frac{1}{6}\eta_A\eta_B \eta_C \epsilon^{ABCD}\psi_{D}^{-} + \eta_1\eta_2 \eta_3\eta_4 A_- \ ,
\end{equation}
where now the component fields and Grassmann variables transform in $SU(4)\times U(1) \sim SO(6)\times SO(2) \subset SO(8)$. The Grassmann variables are built out of the corresponding $SU(2)$ variables in the two Chern-Simons theories: $\eta_A=(\eta_\alpha, \tilde \eta_{\tilde \alpha})$. The ``vector'' fields can be taken to be $A_{\pm} = \phi^0\pm i\phi^7$, and $\phi^0$ is the dualized DBI photon. The fermions can be identified with the DBI fermions as $\psi_{+}^A = \chi^A+i\chi^{A+4}$ and $\psi^{-}_A = \chi_A-i\chi_{A+4}$, where $\chi_{\cal A}=(\chi_{A},\chi_{A+4})$ are the on-shell degrees of freedom carried by the Majorana-Weyl fermions, with ${\cal A}=0,\ldots, 7$ being the $SO(8)$ little group index. Finally, the six scalars $\phi^{AB}$ can be matched to the DBI scalars $\phi^{i=1,\ldots,6}$.

We checked up to multiplicity six that the double copy agrees with the amplitudes obtained from the ${\cal N}=8$ DBI theory, and similarly agrees with the alternative double copy $({\cal N}=8~{\rm SYM}) \otimes {\rm NLSM}$ that involves maximally supersymmetric Yang-Mills theory.  Furthermore, the relationship between the coupling $\kappa$ in \eqn{doubleCopy} and the DBI coupling is as simple as $\alpha^2=\kappa/2$.

\section{Conclusions}

In this paper, we showed that pure Chern-Simons theory obeys the fully off-shell version of color-kinematics duality. We observe that the standard Lorenz-gauge Feynman rules have the property that the kinematic Jacobi relations are automatic both at tree and loop level. Furthermore the duality extends smoothly to the BRST ghost sector of Chern-Simons theory, implying that the complete quantum theory obeys color-kinematics duality. 

Employing a superfield formulation \citep{Axelrod:1991vq} for the entire field content, including Faddeev-Popov  ghosts, we showed that the superspace Feynman rules for the theory can be interpreted as nested Poisson brackets defined by a second-order differential operator, in a similar fashion to the recent Yang-Mills construction in ref.~\cite{Ben-Shahar:2021doh}. 
The computation of BCJ numerators for any Feynman diagram, including loops, is thus straightforward. We furthermore identify the corresponding kinematic algebra of pure Chern-Simons theory as the algebra of volume-preserving diffeomorphisms. 

While pure Chern-Simons theory has no on-shell amplitudes, the observation that the off-shell Feynman rules obey color-kinematics duality allows one to double-copy Chern-Simons theory at the Lagrangian level. We do this for several theories, including when one copy is Yang-Mills  theory or a higher derivative $(DF)^2$ theory. These theories can be double copied with Chern-Simons theory, since it is sufficient to have one theory that manifests color-kinematics duality. 
The resulting double-copy theories typically end up having gauge-fixed Lagrangians, and in some cases their kinetic terms are non-local, implying that further work is required for identifying the theories. The double-copy theories also inherit the vanishing of the on-shell amplitudes from the Chern-Simons parent theory, meaning that we cannot compare to gauge-invariant quantities of the candidate theories. Nevertheless, we identify an interesting candidate theory for the (Chern-Simons)$\otimes (DF)^2$ double copy, which we expect to be related to the Chern-Simons-gravity theory in 3D~\cite{Deser1982}.

It would be interesting to explore the Chern-Simons double-copy theories further, and to see if they inherit any useful topological properties from their parent theory. Furthermore, investigating the implications for Wilson-loop (or worldline) observables (see e.g.~\cite{Alfonsi:2020lub,Alawadhi:2021uie,Bastianelli:2021rbt,Shi:2021qsb}) both in Chern-Simons theory, and in the double-copy theories, may be a promising future direction. It would also be interesting to apply the techniques of this paper to the Feynman rules of the Yang-Mills pure-spinor formulation considered in ref.~\citep{Ben-Shahar:2021doh}, which exhibited color-kinematics duality via a Chern-Simons-like Lagrangian. 

For adjoint Chern-Simons matter, we found that there exist theories with doublet scalars, doublet fermions, or supersymmetric ${\cal N}=4$ matter, that obey the BCJ relations. The choice of using the adjoint representation for the matter had the unintended consequence that the scalars and fermions need to obey odd and even statistics, respectively. After combining the scalar and fermion sectors, the BCJ relations determined the non-minimal couplings uniquely to give a theory that exhibits $\mathcal{N}=4$ supersymmetry. While the matter statistics is unusual, we observed that the color-ordered tree amplitudes in this theory precisely match those of the $\mathcal{N}=4$ Gaiotto-Witten theory, which is a supersymmetry truncation of the $\mathcal{N}=6$ ABJM theory. Both of these theories use bi-fundamental matter with standard statistical behavior.  We note that the $\mathcal{N}=4$ Chern-Simons-matter theory has the maximal supersymmetry permitted by the standard BCJ amplitude relations, hence our considerations cannot be uplifted to the $\mathcal{N}=6$ ABJM theory. Finally, the opposite-statistics states in the Chern-Simons-matter theory become standard-statistics states when feeding them through the double copy. Hence the double copy of $\mathcal{N}=4$ Chern-Simons-matter theory with itself gives a well-behaved theory, which we identify with the maximally supersymmetric $\mathcal{N}=8$ DBI theory.

\section*{Aknowledgements}
We would like to thank Lucile Cangemi, Marco Chiodaroli, Simon Ekhammar, Lucia Garozzo, Max Guillen, Joe Minahan and Oliver Schlotterer for enlightening discussions related to this work.
This research is supported in part by the Knut and Alice Wallenberg Foundation under grants KAW 2018.0116 ({\it From Scattering Amplitudes to Gravitational Waves}) and KAW 2018.0162 ({\it Exploring a Web of Gravitational Theories through Gauge-Theory Methods}), the Swedish Research Council under grant 621-2014-5722, and the Ragnar S\"{o}derberg Foundation (Swedish Foundations' Starting Grant). Computational resources (project SNIC 2019/3-645) were provided by the Swedish National Infrastructure for Computing (SNIC) at UPPMAX, partially funded by the Swedish Research Council through grant no. 2018-05973.

\clearpage

\appendix
\section{Conventions and Spinors in 3D}\label{spinorconventions}
We use a mostly minus metric $\eta_{\mu\nu} = \textrm{diag}(1,-1,-1)$, and define 3D gamma matrices as
\begin{equation}
\gamma^{0} = 
\begin{pmatrix}
0 && -1 \\ 1 && 0 
\end{pmatrix} \ ,
\hspace{1cm}
\gamma^{1} = 
\begin{pmatrix}
-1 && 0 \\ 0 && 1 
\end{pmatrix} \ ,
\hspace{1cm}
\gamma^{2} = 
\begin{pmatrix}
0 && 1 \\ 1 && 0 
\end{pmatrix} \ ,
\end{equation}
where the $SL(2,\mathbb{R})\sim SO(1,2)$ indices are distributed as $(\gamma^\mu)_\alpha^{\ \beta}$. The matrices obey the Clifford algebra
\begin{equation}
\{\gamma^\mu , \gamma^\nu\} = -2 \eta^{\mu\nu} \ .
\end{equation}
Spinor indices can be lowered (and raised) with the $SL(2,\mathbb{R})$ metric \begin{equation}
\epsilon_{\alpha\beta} = \begin{pmatrix}
0 && -1 \\ 1 && 0
\end{pmatrix} \ .
\end{equation}
The angle-bracket spinors are defined by the relations
\begin{equation}
\langle p|^\alpha =\epsilon^{\alpha\beta}|p\rangle_\beta \ , \hspace{1cm}
|p \rangle_\alpha\langle p|^\beta  = - p^\mu (\gamma_\mu)_\alpha^{\ \beta} =- \slashed{p}_\alpha^{\ \beta}  \ .
\end{equation}
With this definition, we have that
\begin{equation}
s_{ij} = (p_i+p_j)^2 = \langle i j\rangle ^2 \ .
\end{equation}
In addition, the gamma matrices obey
\begin{equation}
(\gamma^\mu)_{\alpha\beta}(\gamma_\mu)^{\gamma\delta} = -\delta^\gamma_\alpha\delta^\delta_\beta - \delta^\gamma_\beta\delta^\delta_\alpha \ ,
\end{equation}
or, alternatively, 
\begin{equation}
(\gamma^\mu)_\alpha^{\ \beta} (\gamma_\mu)_\gamma^{\ \delta} = \delta_\alpha^\beta \delta_\gamma^\delta - 2 \delta^\beta_\gamma \delta ^\delta_\alpha \ ,
\end{equation}
which implies the following relation when sandwiched with four spinors:
\begin{equation}
\langle 1| \gamma^\mu |2\rangle\langle 3| \gamma_\mu |4\rangle = \langle 12 \rangle \langle 34 \rangle + 2 \langle 1 4\rangle \langle 2 3 \rangle \ .
\end{equation}
Since the 3D Levi-Civita symbol appears frequently, it is useful to note that
\begin{equation}
\epsilon_{\mu\nu\rho}p^\mu q^\nu r^\rho = -\frac{1}{2}\textrm{Tr}(pqr)=\frac{1}{2}\langle pq\rangle \langle qr\rangle \langle rp\rangle \ ,
\end{equation}
and
\begin{equation}
\langle p | \gamma^\mu |q\rangle \epsilon_{\mu \nu\rho}(p+q)^\nu = \langle p q\rangle (p-q)_\rho \ .
\end{equation}

\section{Feynman rules}\label{Feynmanrules}
To write down the Feynman rules, we use an $SU(N_c)$ gauge group, where the generators and structure constants satisfy $\textrm{Tr}(\tilde{T}^a\tilde{T}^b) = \delta^{ab}$ and $[\tilde{T}^a,\tilde{T}^b]=\tilde{f}^{abc}T^c$. In terms of the non-tilde structure constants, we have $
\tilde{f}^{abc} = i\sqrt{2}f^{abc} $, and the covariant derivative is $(D_\mu \phi)^a= \partial_\mu\phi^a + ig\frac{1}{\sqrt{2}}\tilde{f}^{abc}A^b_\mu \phi^c$.  With these conventions, we again give the ${\cal N}=4$ Chern-Simons-matter Lagrangian (\ref{Neq4Lagr}), 
\begin{eqnarray}\label{fullmatterlagrangian}
\mathcal{L}_{\mathcal{N}=4} &=& \epsilon_{\mu\nu\rho}\left(\frac{1}{2}
 A^{a \mu} \partial^\nu A^{a \rho} + \frac{ig}{6\sqrt{2}}\tilde{f}^{abc}A^{a\mu} A^{b\nu} A^{c\rho}\right)
 \nonumber \\
&&\null +
\partial_\mu \bar{\phi}^a\partial^\mu \phi^a +
\frac{i g}{\sqrt{2}}\tilde{f}^{abc}A_\mu^b(\partial_\mu\bar{\phi}^a\phi^c+\bar{\phi}^c\partial^\mu\phi^a )
-
\frac{g^2}{2}\tilde{f}^{abx}\tilde{f}^{xcd}\bar{\phi}^a A^b\cdot A^c \phi^d \nonumber \\
&&\null +
i\bar{\psi}^a\slashed{\partial}\psi^a 
-
\frac{g}{\sqrt{2}}\tilde{f}^{abc}\bar{\psi}^a \slashed{A}^b\psi^c 
-
\frac{i g^2}{2}\bar{\psi}^a \psi^b \bar{\phi}^c\phi^d (\tilde{f}^{acx}\tilde{f}^{xbd} 
+
\tilde{f}^{adx}\tilde{f}^{xbc})\nonumber \\
&&\null -
\frac{ g^4}{4}\phi^{a}\bar{\phi}^{b}\bar{\phi}^{c}\phi^{d}\phi^{e}\bar{\phi}^{h}{f}^{a b x}{f}^{x c y}{f}^{y d z}{f}^{z e h}
 \ .
\end{eqnarray}
The corresponding color-ordered Feynman rules up to multiplicity four are\footnote{Note that the indices on the fermions $\psi^{\alpha}$ are $SL(2,\mathbb{R})$ spinor indices, not R-symmetry indices.}
\begin{equation}
\begin{tikzpicture}[baseline={(0, -0.1cm)}]
\draw[thick] (-0.5,0) -- (0.5,0);
\node at (-0.7,0) {$\bar{\psi}$};
\node at (0.7,0) {$\psi$};
\end{tikzpicture}
=
i\frac{\slashed{p}}{p^2} \ ,
\hspace{1cm}
\begin{tikzpicture}[baseline={(0, -0.1cm)}]
\draw[thick,dashed] (-0.5,0) -- (0.5,0);
\node at (-0.7,0) {$\phi$};
\node at (0.7,0) {$\bar{\phi}$};
\end{tikzpicture}
=
\frac{i}{p^2} \ ,
\hspace{1cm}
\begin{tikzpicture}[baseline={(0, -0.1cm)}]
\draw[thick,snake it] (-0.5,0) -- (0.5,0);
\node at (-0.7,0) {$A_\mu$};
\node at (0.7,0) {$A_\rho$};
\end{tikzpicture}
=
-\frac{\epsilon_{\mu\nu\rho}p^\nu}{p^2}\ ,
\end{equation}

\begin{eqnarray}
\begin{tikzpicture}[scale = 1, baseline={(0, -0.1cm)}]
\draw[thick,snake it] (0,0.75) -- (0,0);
\draw[thick,snake it] (0,0) -- (0.87*0.75,-0.5*0.75);
\draw[thick,snake it] (0,0) -- (-0.87*0.75,-0.5*0.75);
\node at (0,1) {$\mu$};
\node at (0.87,-0.5) {$\nu$};
\node at (-0.87,-0.5) {$\rho$};
\end{tikzpicture}
\!\!\!
=-\frac{\epsilon^{\mu\nu\rho} }{\sqrt{2}} \ ,
\hspace{0.3cm}
\begin{tikzpicture}[scale = 1, baseline={(0, -0.1cm)}]
\draw[thick,snake it] (0,0.75) -- (0,0);
\draw[thick,dashed] (0,0) -- (0.87*0.75,-0.5*0.75);
\draw[thick,dashed] (0,0) -- (-0.87*0.75,-0.5*0.75);
\node at (0,1) {$\mu$};
\node at (0.87,-0.5) {$\bar{\phi}_2$};
\node at (-0.87,-0.5) {$\phi_1$};
\end{tikzpicture}
\!\!\!
=i\frac{(p_1-p_2)^\mu}{\sqrt{2}} \ ,
\hspace{0.3cm}
\begin{tikzpicture}[baseline={(0, -0.1cm)}]
\draw[thick,snake it] (0,0.75) -- (0,0);
\draw[thick] (0,0) -- (0.87*0.75,-0.5*0.75);
\draw[thick] (0,0) -- (-0.87*0.75,-0.5*0.75);
\node at (0,1) {$\mu$};
\node at (0.87,-0.5) {$\psi_\beta$};
\node at (-0.87,-0.5) {$\bar{\psi}^\alpha$};
\end{tikzpicture}
\!\!\!
=i\frac{ (\gamma^\mu)_\alpha^{\ \beta}}{\sqrt{2}} \ , ~~~
\end{eqnarray}

\begin{eqnarray}
&&\begin{tikzpicture}[baseline={(0, -0.1cm)}]
\draw[thick,snake it] (-0.5,0.5) -- (0,0);
\draw[thick,snake it] (-0.5,-0.5) -- (0,0);
\draw[thick,dashed] (0.5,0.5) -- (0,0);
\draw[thick,dashed] (0.5,-0.5) -- (0,0);
\node at (-0.5*1.25,-0.55*1.25) {$A_\mu$};
\node at (-0.5*1.25,0.55*1.25) {$A_\nu$};
\node at (0.5*1.25,0.55*1.25) {$\phi$};
\node at (0.5*1.25,-0.55*1.25) {$\bar{\phi}$};
\end{tikzpicture}
=-\frac{i}{2}\eta^{\mu\nu} \ ,
\hspace{0.3cm}
\begin{tikzpicture}[baseline={(0, -0.1cm)}]
\draw[thick,dashed] (-0.5,0.5) -- (0.5,-0.5);
\draw[thick,snake it] (-0.5,-0.5) -- (0.5,0.5);
\node at (-0.5*1.25,-0.55*1.25) {$A_\mu$};
\node at (-0.5*1.25,0.55*1.25) {$\phi$};
\node at (0.5*1.25,0.55*1.25) {$A_\nu$};
\node at (0.5*1.25,-0.55*1.25) {$\bar{\phi}$};
\end{tikzpicture}
=-i\eta^{\mu\nu} \ ,
\\
&&
\begin{tikzpicture}[baseline={(0, -0.1cm)}]
\draw[thick,dashed] (-0.5,0.5) -- (0,0);
\draw[thick] (-0.5,-0.5) -- (0,0);
\draw[thick,dashed] (0.5,0.5) -- (0,0);
\draw[thick] (0.5,-0.5) -- (0,0);
\node at (-0.5*1.25,-0.55*1.25) {$\psi_\beta$};
\node at (-0.5*1.25,0.55*1.25) {$\phi$};
\node at (0.5*1.25,0.55*1.25) {$\bar\phi$};
\node at (0.5*1.25,-0.55*1.25) {$\bar{\psi}^\alpha$};
\end{tikzpicture} 
=
-\frac{1}{2}\delta^\alpha_\beta \ ,
\hspace{0.3cm}
\begin{tikzpicture}[baseline={(0, -0.1cm)}]
\draw[thick,dashed] (-0.5,0.5) -- (0.5,-0.5);
\draw[thick] (-0.5,-0.5) -- (0.5,0.5);
\node at (-0.5*1.25,-0.55*1.25) {$\psi^\alpha$};
\node at (-0.5*1.25,0.55*1.25) {$\phi$};
\node at (0.5*1.25,0.55*1.25) {$\bar{\psi}_\beta$};
\node at (0.5*1.25,-0.55*1.25) {$\bar{\phi}$};
\end{tikzpicture}
=\delta_\alpha^\beta \ .
\end{eqnarray}
It is easy to restore the free coefficients $\alpha$ and $\beta$ introduced in section~\ref{Neq2}, in which case the last two scalar-fermion rules would be multiplied by $\alpha$ and $(\alpha+\beta)/2$, respectively. The rules here were obtained for $\alpha=1$ and $\beta=1$. 
The color-ordered six-point scalar interactions are
\begin{eqnarray}
\begin{tikzpicture}[baseline={(0, -0.1cm)}]
\draw[thick,dashed] (0,0.75) -- (0,0);\node at (0,1) {$\bar{\phi}$};
\draw[thick,dashed] (0,-0.75) -- (0,0);\node at (0,-1) {${\phi}$};
\draw[thick,dashed] (0,0) -- (0.87*0.75,-0.5*0.75);\node at (0.87*0.9,-0.5*0.9) {${\phi}$};
\draw[thick,dashed] (0,0) -- (0.87*0.75,0.5*0.75);\node at (0.87*0.9,0.5*0.9) {$\bar{\phi}$};
\draw[thick,dashed] (0,0) -- (-0.87*0.75,-0.5*0.75);\node at (-0.87*0.9,-0.5*0.9) {${\phi}$};
\draw[thick,dashed] (0,0) -- (-0.87*0.75,0.5*0.75);\node at (-0.87*0.9,0.5*0.9) {$\bar{\phi}$};
\end{tikzpicture}
=
-\frac{i}{2} \ ,
\ \ \ \ \ \
\begin{tikzpicture}[baseline={(0, -0.1cm)}]
\draw[thick,dashed] (0,0.75) -- (0,0);\node at (0,1) {$\bar{\phi}$};
\draw[thick,dashed] (0,-0.75) -- (0,0);\node at (0,-1) {${\phi}$};
\draw[thick,dashed] (0,0) -- (0.87*0.75,-0.5*0.75);\node at (0.87*0.9,-0.5*0.9) {${\phi}$};
\draw[thick,dashed] (0,0) -- (0.87*0.75,0.5*0.75);\node at (0.87*0.9,0.5*0.9) {$\bar{\phi}$};
\draw[thick,dashed] (0,0) -- (-0.87*0.75,-0.5*0.75);\node at (-0.87*0.9,-0.5*0.9) {$\bar{\phi}$};
\draw[thick,dashed] (0,0) -- (-0.87*0.75,0.5*0.75);\node at (-0.87*0.9,0.5*0.9) {${\phi}$};
\end{tikzpicture}
=
\frac{i}{4}\ ,
\ \ \ \ \ \ 
\begin{tikzpicture}[baseline={(0, -0.1cm)}]
\draw[thick,dashed] (0,0.75) -- (0,0);\node at (0,1) {${\phi}$};
\draw[thick,dashed] (0,-0.75) -- (0,0);\node at (0,-1) {$\bar{\phi}$};
\draw[thick,dashed] (0,0) -- (0.87*0.75,-0.5*0.75);\node at (0.87*0.9,-0.5*0.9) {${\phi}$};
\draw[thick,dashed] (0,0) -- (0.87*0.75,0.5*0.75);\node at (0.87*0.9,0.5*0.9) {$\bar{\phi}$};
\draw[thick,dashed] (0,0) -- (-0.87*0.75,-0.5*0.75);\node at (-0.87*0.9,-0.5*0.9) {${\phi}$};
\draw[thick,dashed] (0,0) -- (-0.87*0.75,0.5*0.75);\node at (-0.87*0.9,0.5*0.9) {$\bar{\phi}$};
\end{tikzpicture}
=
0 \ ,
\end{eqnarray}
where it is important to keep in mind that the sign of the vertices flips under cyclic permutations, because of the anti-commuting nature of the scalars.  The fudge factor $\lambda=1$ can be restored by multiplication if needed.

\newpage
\bibliographystyle{JHEP}
\bibliography{bib}

\end{document}